\documentclass[11pt]{article}
\usepackage{myarticle}

\usepackage{centernot} 

\usepackage[frenchb, english]{babel}
\usepackage[T1]{fontenc}

\usepackage[utf8]{inputenc}  

\usepackage{alltt}
\usepackage{fancyvrb}

\usepackage[normalem]{ulem}

\usepackage{xcolor, colortbl, graphicx}
\definecolor{lightred}{RGB}{250,240,240}

\usepackage{subfig}

\usepackage[nottoc, notlof, notlot]{tocbibind}

\usepackage{caption}
\usepackage{amsfonts, amssymb, amsthm} 
\usepackage[tbtags]{mathtools}   
\numberwithin{equation}{section} 
\allowdisplaybreaks              

\usepackage{stmaryrd}
\usepackage{bbold}     

\renewcommand{\imath}{i} 
\renewcommand{\Re}[1]{\operatorname{Re}(#1)} 
\renewcommand{\Im}[1]{\operatorname{Im}(#1)} 

\newcommand{\hc}{\text{h.c.}}

\makeatletter
\def\resizemath#1{%
	\begingroup\makeatletter#1\check@mathfonts
	\def\maketag@@@##1{\hbox{\m@th\normalsize\normalfont##1}}\aftergroup
}
\def\endresizemath{\endgroup}
\makeatother

\usepackage{hyperref}
\hypersetup{
	backref=true,            
	pagebackref=true,        
	breaklinks=true,         
	hyperindex=true,         
	hyperfootnotes=true,     
	colorlinks=true,         
	linkcolor=red,           
	linktoc=page,            
	citecolor=blue, 
	filecolor=magenta,       
	urlcolor=blue,           
	runcolor=filecolor,      
	menucolor=red,           
	pdfdisplaydoctitle=true, 
	bookmarks=true,          
	bookmarksopen=false,     
	unicode=true,            
	pdftitle={Effective Higgs Lagrangian and Constraints on Higgs Couplings}, 
	pdfauthor={Hermès BÉLUSCA - MAÏTO}, 
	pdfwindowui=true,        
	pdftoolbar=true,         
	pdfmenubar=true,         
	pdffitwindow=false,      
	pdfstartview={Fit},      
	pdfnewwindow=true        
}

\usepackage[referable]{threeparttablex}

\usepackage{siunitx}
\usepackage{tabularx}
\newcolumntype{C}{>{\centering\arraybackslash}X}

\usepackage{multirow, bigdelim}

\usepackage{abstract}

\usepackage[]{authblk}

\usepackage{collref}

\newcommand{\sectionnn}[1]{\section*{#1}\addcontentsline{toc}{section}{#1}}

\newcommand{\email}[1]{\href{mailto:#1}{\protect\nolinkurl{#1}}}

\title{\vfill \bfseries \textsc{Effective Higgs Lagrangian and \\ Constraints on Higgs Couplings}}
\author{Hermès~\textsc{Bélusca--Maïto}\footnote{\email{hermes.belusca@th.u-psud.fr}}}
\affil{Laboratoire de Physique Théorique, CNRS -- UMR 8627, \\ Bât. 210, Université Paris-Sud XI, Université Paris-Saclay, F-91405 \textsc{Orsay} Cedex, France}
\preprint{LPT-Orsay-14-22}

\date{}

\begin{document}

\addtolength{\belowdisplayskip}{-2pt}
\addtolength{\belowdisplayshortskip}{-2pt}


	\pagestyle{empty}
	\maketitle
	\thispagestyle{empty}
	\vfill
	\begin{abstract}
		\par Probing the properties of the discovered Higgs boson may tell us whether or not it is the same particle as the one predicted by the Standard Model. To this aim we parametrize deviations of the Higgs couplings to matter from the Standard Model by using the Higgs Effective Field Theory framework. Starting with a general dimension-6 effective Lagrangian including both CP-even and CP-odd operators, and requiring that the operators do not introduce power divergences in the oblique parameters, we reduce the number of independent effective couplings of the theory. This framework is then used to put updated constraints on the effective couplings, using the latest Higgs rates data from the Run-I of the ATLAS and CMS experiments, and electroweak precision data from LEP, SLC and Tevatron. We show that the current data is able to significantly constrain the CP-even and some CP-odd operators of the effective Lagrangian.
	\end{abstract}
	\vfill

\newpage
	\tableofcontents

\newpage
\pagestyle{plain}
\setcounter{page}{1}
\setcounter{footnote}{0}

\sectionnn{Introduction}
	\par The recent experimental confirmation by the ATLAS and CMS experiments~\cite{Aad:2012tfa,Chatrchyan:2012ufa} of the existence of a scalar particle with mass $m_h$ of about $125$~GeV and production cross-sections and decay rates compatible with those of the Standard Model (SM) Higgs boson, triggered many studies of its properties and put constraints on the SM and on some Beyond-the-Standard Model (BSM) theories. However many questions remain open concerning the fundamental nature of this boson and the dynamics of the electroweak symmetry breaking (EWSB). If we want to know whether this discovered scalar particle is \emph{the} SM or \emph{a} SM-like Higgs boson as predicted by some BSM models, probing its properties with precision becomes mandatory. Two main strategies are available for studying its properties and their possible deviations from their SM predictions, by either performing a study in the context of a specified model, or, as employed in this paper, using a model-independent approach with an effective theory framework. The Wilson coefficients of the effective operators parametrize in a continuous way the possible deviations of the Higgs couplings from their SM values.

	\par The purpose of this paper is twofold: a phenomenological Higgs Lagrangian is derived by using the effective dimension-6 operators and conditions from the oblique $S$, $T$, $U$ parameters, then the constraints on the phenomenological Lagrangian parameters are obtained by performing a global fit using a combination of the latest Higgs signal rates from the Run-I of ATLAS and CMS experiments, together with electroweak (EW) precision measurements from LEP, SLC and Tevatron. As such, this paper is a continuation of numerous previous studies, amongst which some that mainly targeted only CP-conserving couplings~\cite{Carmi:2012zd,Carmi:2012in,Falkowski:2013dza,Giardino:2013bma,Ellis:2013lra,Belanger:2013xza,Ellis:2014dva,Ellis:2014jta,Corbett:2015ksa} or others considering CP-conserving as well as CP-violating couplings~\cite{Freitas:2012kw,Djouadi:2013qya,Dumont:2013wma,Brod:2013cka,Dwivedi:2015nta}, using either a simplified phenomenological Higgs Lagrangian or the effective dimension-6 Lagrangian before EWSB in a certain choice of operator basis. Albeit LHC data strongly suggests that the observed particle is in excellent agreement with the SM predictions for the Higgs boson and it indeed possesses the required quantum numbers for a scalar particle~\cite{Aad:2013xqa,Chatrchyan:2012jja,Lenzi:2013CernTalk}, we will nevertheless consider the possibility that this particle may have both CP-conserving and CP-violating couplings to bosons and fermions. We will observe that the CP-even and some of the CP-odd couplings are significantly constrained using the existing experimental data.

	\par This paper is organised as follows. In the first section we review the effective Lagrangian in the linear realization, built from $SU(3)_C \times SU(2)_L \times U(1)_Y$-invariant dimension-6 operators where the Higgs field is embedded in a weak $SU(2)_L$ doublet~$H$. Because the effective Lagrangian contains a too large number of parameters for practical phenomenological purposes, this number is reduced in the next section by imposing extra relations amongst these parameters. Those relations are of two kinds: the first ones come from the nature of the dimension-6 Lagrangian used as the starting point, and the second ones are needed to remove large power-divergent contributions to the oblique parameters. After imposing these constraints a total of 7 independent effective couplings in the CP-conserving sector and 6 independent ones in the CP-violating sector is obtained. Finally we compute the theoretical Higgs signal strengths in the effective framework and fit them to the ones coming from the latest Run-I Higgs data from ATLAS and CMS in various production and decay channels, combined with the EW precision measurements, and we obtain constraints for the CP-even and CP-odd couplings which are then discussed. The details of computations are relegated to the appendices.

	\section{Dimension-6 EFT and Phenomenological Higgs Lagrangian}
	\label{sec:EffLagr}
		\par Several assumptions on the physics of the involved Higgs boson are made, amongst those the first and critical one being that new-physics (NP) degrees of freedom should reside at an energy scale much higher than the EW scale. The NP fields are integrated-out and give rise, at lower energies, to higher dimension effective non-renormalizable operators in the expansion of the effective Lagrangian, inducing deviations of the leading-order (LO) Higgs couplings from their SM values. In this work we also require baryon and lepton (BL) numbers conservation and the absence of any source of flavour violation.

		\par We assume that the Higgs boson $h$ is part of the Higgs field $H$ which transforms in the $(\textbf{1},\textbf{2},\frac{1}{2})$ representation of the Standard Model $SU(3)_C \times SU(2)_L \times U(1)_Y$ gauge group and acquires an expectation value~$v$. The effective Lagrangian is then expanded:
		\begin{equation}
			\mathcal{L}_{eff} = \mathcal{L}_\text{SM} + \mathcal{L}_{D=5} + \mathcal{L}_{D=6} + \dots
		\end{equation}
		where each part consists of gauge-invariant local operators of canonical dimension $D$ that are made uniquely of SM fields. The leading term in this expansion is the SM Lagrangian which contains operators up to dimension 4 (see Appendix~\ref{subsec:SMLagr} for notations). At the level of dimension-5 operators there is only one respecting the SM gauge symmetry (Weinberg operator) which gives masses to neutrinos after EWSB and does not have any impact on the Higgs phenomenology; however it violates lepton number conservation so it is removed from our study. The part of interest consists in the dimension-6 operators. Requiring BL numbers conservation, it is known~\cite{Grzadkowski:2010es} there are 59 operators\footnote{Although their original list was known since the 1980's~\cite{Buchmuller:1985jz} (assuming BL conservation, 80 operators were found), it was pointed out by many analyses that some of them were redundant, and a complete minimal list of 59 operators was finally given by Grzadkowski {\it et~al.}~\cite{Grzadkowski:2010es} in 2010.} that form a basis of dimension-6 operators for a single generation\footnote{This is the number of operators when considering only one generation; otherwise their number increase to 2499, see the review~\cite{Alonso:2013hga} for the details of counting.}. A choice of operator basis needs to be made because these operators can be redefined into other ones using equations of motion. In this paper we choose the SILH basis\footnote{Other choices of bases are of course possible, amongst which the HISZ basis~\cite{Hagiwara:1993ck,Corbett:2015ksa} and the Warsaw basis~\cite{Grzadkowski:2010es}. After the first version of this article was issued, motivated by experimental Higgs analyses, \cite{Gupta:2014rxa} introduced the so-called "Higgs primary couplings" (see also~\cite{Pomarol:2014dya}); the LHC Higgs Cross Section Working Group 2, for the same reasons, is developing a EFT framework~\cite{HXSWGbasis} in which LHC Higgs observables could easily be linked to the Wilson coefficients; as such, they also assume a linear-realized EFT with a $SU(2)$ Higgs doublet. The connection with experiments is done by the choice of unitary gauge and writing all the Lagrangian in terms of the physical fields --~bosons, fermions and Higgs field~--, this is the so-called "Higgs basis".} employed by Contino~{\it et~al.}~\cite{Contino:2013kra}.

		\par In this framework the dimension-6 part of the effective Lagrangian can be written as:
		\begin{equation}
			\mathcal{L}_{D=6} = \mathcal{L}_{CPC} + \mathcal{L}_{CPV}
		\end{equation}
		where the CP-conserving part is given by: $\mathcal{L}_{CPC} = \Delta\mathcal{L}_{SILH} + \Delta\mathcal{L}_{F_1} + \Delta\mathcal{L}_{F_2} + \Delta\mathcal{L}_{4F} + \Delta\mathcal{L}_{Gauge}$ using the notations of~\cite{Contino:2013kra} (see also their Equations 2.1, 2.2, 2.3, 2.4 and 2.6): $\Delta\mathcal{L}_{SILH}$ is the Strongly-Interacting Light Higgs doublet Lagrangian (SILH) first introduced by Giudice {\it et~al.}~\cite{Giudice:2007fh}, $\Delta\mathcal{L}_{F_1}$ contains the 2-fermion vertex operators and $\Delta\mathcal{L}_{F_2}$ contains the 2-fermion dipole operators. $\Delta\mathcal{L}_{4F}$ is constituted of twenty-two 4-fermion baryon-number-conserving operators while $\Delta\mathcal{L}_{Gauge}$ is made of gauge-boson self-interaction operators, which affect the gauge-boson propagators and self-interactions but they do not have any effect on Higgs physics~\cite{Contino:2013kra}. The CP-violating part $\mathcal{L}_{CPV}$ contains all the possible dimension-6 CP-odd operators (Equation~C.96 of~\cite{Contino:2013kra}).

		\par Our first selection of operators is motivated given the current sensitivity of the LHC experiments: operators of dimension greater than 6 will not be relevant. EW precision measurements from LEP strongly constrain the couplings of SM fermions to EW gauge bosons, which are modified in the presence of the 2-fermion operators $\Delta\mathcal{L}_{F_1}$, so that there is not much room to affect LHC Higgs phenomenology. The same remark also applies for the 2-fermion dipole operators $\Delta\mathcal{L}_{F_2}$ which contribute to electric and magnetic dipole moments (EDM and MDM), and also contribute to the 3-body Higgs decay: they are further suppressed and thus can be ignored in the present analysis. The $\Delta\mathcal{L}_{F_4}$ and $\Delta\mathcal{L}_{Gauge}$ parts of the Lagrangian are ignored because they do not involve any Higgs boson; the gauge part modifying only the triple and quartic gauge boson couplings and the oblique parameters. The CP-violating part $\mathcal{L}_{CPV}$ contains also gauge-boson self-interaction operators that are not taken into account here for the same reasons as for $\Delta\mathcal{L}_{Gauge}$. Finally the SILH Lagrangian contains a Higgs self-interaction operator of the form $(H^\dagger H)^3$ that is also removed in this work because current experimental precision is not sensitive enough to modifications of the Higgs self-couplings.

		\par Considering all of these remarks the relevant dimension-6 Lagrangian can be rewritten as:
		\begin{align}
			\begin{split}
			\label{eq:L_cpc}
			\mathcal{L}_{CPC} = & \frac{\bar{c}_H}{2v^2} \partial^\mu (H^\dagger H) \partial_\mu (H^\dagger H) + \frac{\bar{c}_T}{2v^2} \left(H^\dagger \overleftrightarrow{D^\mu} H\right) \left(H^\dagger \overleftrightarrow{D_\mu} H\right) \\
				&+ \frac{H^\dagger H}{v^2} \left( \bar{c}_u y_u \overline{q_L} H^c u_R + \bar{c}_d y_d \overline{q_L} H d_R + \bar{c}_l y_l \overline{L_L} H l_R + \hc \right) \\
				&+ \frac{\imath\,\bar{c}_W~g}{2 m_W^2} \left(H^\dagger \sigma^i \overleftrightarrow{D^\mu} H\right) (D^\nu W_{\mu\nu})^i + \frac{\imath\,\bar{c}_B~g'}{2 m_W^2} \left(H^\dagger \overleftrightarrow{D^\mu} H\right) (\partial^\nu B_{\mu\nu}) \\
				&+ \frac{\imath\,\bar{c}_{HW}~g}{m_W^2} (D^\mu H)^\dagger \sigma^i (D^\nu H) W^i_{\mu\nu} + \frac{\imath\,\bar{c}_{HB}~g'}{m_W^2} (D^\mu H)^\dagger (D^\nu H) B_{\mu\nu} \\
				&+ \frac{\bar{c}_\gamma~g'^2}{m_W^2} (H^\dagger H) B_{\mu\nu} B^{\mu\nu} + \frac{\bar{c}_g~g_S^2}{m_W^2} (H^\dagger H) G^a_{\mu\nu} G^{a{\mu\nu}}
			\end{split} \; ,
			\\[0.5cm]
			\begin{split}
			\label{eq:L_cpv}
			\mathcal{L}_{CPV} = & \frac{\imath\,\widetilde{c}_{HW}~g}{m_W^2} (D^\mu H)^\dagger \sigma^i (D^\nu H) \widetilde{W}^i_{\mu\nu} + \frac{\imath\,\widetilde{c}_{HB}~g'}{m_W^2} (D^\mu H)^\dagger (D^\nu H) \widetilde{B}_{\mu\nu} \\
				&+ \frac{\widetilde{c}_\gamma~g'^2}{m_W^2} (H^\dagger H) B_{\mu\nu} \widetilde{B}^{\mu\nu} + \frac{\widetilde{c}_g~g_S^2}{m_W^2} (H^\dagger H) G^a_{\mu\nu} \widetilde{G}^{a{\mu\nu}}
			\end{split} \; ,
		\end{align}
		where we used the field-strength tensors $F_{\mu\nu}$ and their duals $\widetilde{F}_{\mu\nu} \equiv \frac{1}{2} \epsilon_{\mu\nu\rho\sigma} F^{\rho\sigma}$, and have defined $H^c \equiv \imath\,\sigma_2 H^\star$ and the anti-Hermitian derivative $A^\dagger \overleftrightarrow{D_\mu} B \equiv A^\dagger (D_\mu B) - (D_\mu A)^\dagger B$. $g_S$, $g$ and $g'$ are the $SU(3)_C$, $SU(2)_L$ and $U(1)_Y$ coupling constants, respectively.

		\par It should be noted that we adopt the normalization convention of Contino~{\it et~al.}~\cite{Contino:2013kra} instead of the one from the original SILH article~\cite{Giudice:2007fh} or the usual one, namely: $\frac{\bar{c}_i}{\Lambda_{NP}^2} \mathcal{O}_{D=6}$ where $\Lambda_{NP}$ is the new-physics scale, for the purpose of not making any prior assumptions about its numerical value. In this normalization the scale $\Lambda_{NP}$ is reabsorbed into the Wilson coefficients $\bar{c}_i$, which are then expected to be small if they describe little deviations from the SM. This assumption is so far in agreement with experiments which do not find large deviations from the SM.

		\par The Yukawa-like terms appearing in the second line of Eq.~\ref{eq:L_cpc} are diagonal in the Higgs mass basis because no source of flavour violation is assumed to be present. Also we suppose that the Higgs boson can couple in a different way to up- and down-type quarks, and charged leptons, so that: $\bar{c}_u = \bar{c}_c = \bar{c}_t \equiv \bar{c}_u$, $\bar{c}_d = \bar{c}_s = \bar{c}_b \equiv \bar{c}_d$ and $\bar{c}_e = \bar{c}_\mu = \bar{c}_\tau \equiv \bar{c}_l$. Furthermore these coefficients can be in full generality complex-valued so that their real (resp. imaginary) part give rise to CP-conserving (resp. CP-violating) Higgs couplings to fermions. In the CP-violating piece of the Lagrangian Eq.~\ref{eq:L_cpv}, the field-strengths contractions (last line) give contributions to the $\theta$-terms for $U(1)_Y$ and $SU(3)_C$ so they are expected to be very small.

		\par To finish the build-up of the phenomenological Higgs Lagrangian we place ourselves in unitary gauge and we expand $\mathcal{L}_{eff} = \mathcal{L}_\text{SM} + \mathcal{L}_{D=6}$ after EWSB in the physical Higgs field $h$ around its vacuum expectation value: $H \rightarrow \frac{v}{\sqrt{2}} \left(0; 1+\frac{h}{v}\right)^T$. The $\bar{c}_H$ term of the dimension-6 Lagrangian introduces a finite wave-function renormalization to the Higgs field which needs to be rescaled to bring its kinetic term back into canonical normalization:
		\begin{equation}
		\label{eq:HiggsNormaliz}
			h \rightarrow \frac{h}{\sqrt{1 + \bar{c}_H}} \approx h \left(1 - \frac{\bar{c}_H}{2}\right) \; ,
		\end{equation}
		which effect is to give a universal resizing of all the partial Higgs decay widths, given by the rescaling coefficient of $h$. The dimension-6 Yukawa-like terms shift the mass terms of the fermions (obtained after setting the three Higgs fields of these terms to their vacuum expectation value), and we obtain:
		\begin{equation}\begin{split}
		\label{eq:YukExpand}
			y_f \overline{f_L} H f_R &+ \frac{H^\dagger H}{v^2} \bar{c}_f y_f \overline{f_L} H f_R + \hc \rightarrow m_f^0 \overline{f} f \left(1+\frac{h}{v}\right) + \frac{m_f^0}{2} \overline{f} [\Re{\bar{c}_f} + \imath \gamma_5 \Im{\bar{c}_f}] f \left(1+\frac{h}{v}\right)^3 \\
				&\approx \overline{f} m_f^0 [1 + \frac{\Re{\bar{c}_f} + \imath \gamma_5 \Im{\bar{c}_f}}{2}] f + \frac{h}{v} \overline{f} m_f^0 [1 + \frac{3}{2} (\Re{\bar{c}_f} + \imath \gamma_5 \Im{\bar{c}_f})] f + \mathcal{O}\left(\frac{h^2}{v^2}\right)
		\end{split}\end{equation}
		with $m_f^0 = \frac{y_f v}{\sqrt{2}}$. We choose to reabsorb the correction into a new definition of mass of the fermions\footnote{Alternatively the fermions can be rotated again to put them into their mass basis, therefore diagonalizing their mass term.} and so, by defining $m_f = m_f^0 [1 + \frac{\Re{\bar{c}_f} + \imath \gamma_5 \Im{\bar{c}_f}}{2}]$ we can formally invert this relation and express $m_f^0$ in function of $m_f$. At first order in the $\bar{c}_i$, Eq.~\ref{eq:YukExpand} becomes:
		\begin{equation}
			(\ref{eq:YukExpand}) \rightarrow \overline{f} m_f f + \frac{h}{v} \overline{f} m_f [1 + \Re{\bar{c}_f} + \imath \gamma_5 \Im{\bar{c}_f}] f + \mathcal{O}\left(\frac{h^2}{v^2}\right) \; .
		\end{equation}
		Taking into account Eq.~\ref{eq:HiggsNormaliz} we obtain the values of $c_f$ and $\widetilde{c_f}$ quoted in the dictionary~\ref{eq:cDico}. The $\bar{c}_T$ term in Eq.~\ref{eq:L_cpc} gives a correction to the mass term of the $Z$ boson:
		\begin{equation}
			\frac{m_Z^2}{2} Z_\mu Z^\mu \rightarrow \frac{m_Z^2}{2} (1 - \bar{c}_T) Z_\mu Z^\mu
		\end{equation}
		which should be not normalized if the definition of $\cos{\theta_W} = \frac{m_W}{m_Z}$ is kept. In the next section we will show that $\bar{c}_T$ can be set to zero, hence the $Z$ boson mass correction disappears.

		\par After all these steps, the effective Lagrangian can be written as an expansion in powers of the physical Higgs field $h$: $\mathcal{L}_{eff} = \frac{1}{2}\left(\partial^\mu h\right)^2 - \frac{m_H^2}{2} h^2 + \mathcal{L}_0 + \mathcal{L}_1 + \cdots$; only linear Higgs interactions are kept because LHC experiments are not sensitive to multi-Higgs production up to now. This will however need to be reconsidered for LHC Run-II at higher energies and luminosities. The Higgs-independent part is given by:

		\begin{equation}\begin{split}
		\label{eq:L_0_eff}
			\mathcal{L}_0 = &- \frac{1}{4} F_{\mu\nu} F^{\mu\nu} - \frac{1}{4} Z_{\mu\nu} Z^{\mu\nu} - \frac{1}{4} W^+_{\mu\nu} W^{-{\mu\nu}} - \frac{1}{4} G^a_{\mu\nu} G^{a{\mu\nu}} + \overline{f_L}^i \imath \centernot{D} f_L^i + \overline{f_R}^i \imath \centernot{D} f_R^i \\
				&+ \frac{m_W^2}{2} W^+_{\mu} W^{-{\mu}} + \frac{m_Z^2}{2} (1 - \bar{c}_T) Z_\mu Z^\mu - \sum_{f=u,d,l} m_f \overline{f} f \\
				&+ 2 \bar{c}_\gamma \tan^2{\theta_W} \left(s_w^2 Z_{\mu\nu} Z^{\mu\nu} + c_w^2 \gamma_{\mu\nu} \gamma^{\mu\nu} - 2 s_w c_w Z_{\mu\nu} \gamma^{\mu\nu} \right) + 2 \bar{c}_g \frac{g_S^2}{g^2} G_{\mu\nu} G^{\mu\nu} + \text{CP-Odd} \\
				&+ \bar{c}_B Z^\mu \partial^\nu \left(\tan^2{\theta_W} Z_{\mu\nu} - \tan{\theta_W} \gamma_{\mu\nu}\right) + \text{CP-Odd} \\
				&+ \bar{c}_W \left(\tan{\theta_W} Z^\mu \partial^\nu \gamma_{\mu\nu} + Z^\mu \partial^\nu Z_{\mu\nu} + W^\mu D^\nu W_{\mu\nu}^\dagger + \hc\right) + \text{CP-Odd} \\
				&+ \bar{c}_{HB} \times \text{3-boson} + \bar{c}_{HW} \times \text{3-boson} + \text{CP-Odd}
		\end{split}\end{equation}
		where "CP-Odd" holds for the equivalent CP-odd operators (${V_1}_{\mu\nu}{V_2}^{\mu\nu} \rightarrow {V_1}_{\mu\nu}\widetilde{V_2}^{\mu\nu}$), and "3-boson" holds for operators containing a product of 3 gauge bosons or more, not considered in the rest of this work. The linear part is:
		\begin{equation}
		\label{eq:L_1_eff}
			\begin{split}
			\mathcal{L}_1 = \frac{h}{v} \left[\vphantom{\sum}\right.
			& 2 c_W m_W^2 W_{\mu}^\dagger W^{\mu} + c_Z m_Z^2 Z_{\mu} Z^{\mu} - \sum_{f=u,d,l} m_f \overline{f}\left(c_f + \imath\gamma_5\,\widetilde{c_f}\right)f \\
			& - \frac{1}{2} c_{WW} W_{\mu\nu}^\dagger W^{\mu\nu} - \frac{1}{4} c_{ZZ} Z_{\mu\nu} Z^{\mu\nu} - \frac{1}{4} c_{\gamma\gamma} \gamma_{\mu\nu} \gamma^{\mu\nu} - \frac{1}{2} c_{Z\gamma} \gamma_{\mu\nu} Z^{\mu\nu} + \frac{1}{4} c_{gg} G^a_{\mu\nu} G^{a{\mu\nu}} \\
			& - \frac{1}{2} \widetilde{c}_{WW} W_{\mu\nu}^\dagger \widetilde{W}^{\mu\nu} - \frac{1}{4} \widetilde{c}_{ZZ} Z_{\mu\nu} \widetilde{Z}^{\mu\nu} - \frac{1}{4} \widetilde{c}_{\gamma\gamma} \gamma_{\mu\nu} \widetilde{\gamma}^{\mu\nu} - \frac{1}{2} \widetilde{c}_{Z\gamma} \gamma_{\mu\nu} \widetilde{Z}^{\mu\nu} + \frac{1}{4} \widetilde{c}_{gg} G^a_{\mu\nu} \widetilde{G}^{a{\mu\nu}} \\
			& - (\kappa_{WW} W^\mu D^\nu W_{\mu\nu}^\dagger + \hc) - \kappa_{ZZ} Z^\mu \partial^\nu Z_{\mu\nu} - \kappa_{Z\gamma} Z^\mu \partial^\nu \gamma_{\mu\nu}
			\left.\vphantom{\sum}\right]
			\end{split}
		\end{equation}
		where $s_w = \sin{\theta_W}$ and $c_w = \cos{\theta_W}$ are the sine and cosine of the weak angle. No $\gamma^\mu \partial^\nu \gamma_{\mu\nu}$ or $\gamma^\mu \partial^\nu Z_{\mu\nu}$ terms are present in $\mathcal{L}_1$ because they break $U(1)_{EM}$ symmetry, and also no CP-odd term $\widetilde{\kappa} V^\mu D^\nu \widetilde{V}_{\mu\nu}$ because they cancel~\cite{Contino:2013kra} via the Bianchi identity for the field-strength tensor ${V}_{\mu\nu}$.

		\par The dictionary between the couplings $\bar{c}_i$ and $c_i$ of the Lagrangians~\ref{eq:L_cpc}, \ref{eq:L_cpv} and~\ref{eq:L_1_eff} reads:
		\begin{equation}\begin{alignedat}{3}
		\label{eq:cDico}
			c_W &= 1 - \frac{\bar{c}_H}{2} &\quad&;\quad& c_Z &= 1 - \frac{\bar{c}_H}{2} - \bar{c}_T \\
			c_f &= 1 - \frac{\bar{c}_H}{2} + \Re{\bar{c}_f} &\quad&;\quad& \widetilde{c_f} &= \Im{\bar{c}_f} \quad\text{where}\quad f=u,d,l \\
			c_{WW} &= 4 \bar{c}_{HW} &\quad&;\quad& \widetilde{c}_{WW} &= 4 \widetilde{c}_{HW} \\
			c_{ZZ} &= 4 \left(\bar{c}_{HW} + \frac{s_w^2}{c_w^2} \bar{c}_{HB} - 4 \frac{s_w^4}{c_w^2} \bar{c}_\gamma\right) &\quad&;\quad& \widetilde{c}_{ZZ} &= 4 \left(\widetilde{c}_{HW} + \frac{s_w^2}{c_w^2} \widetilde{c}_{HB} - 4 \frac{s_w^4}{c_w^2} \widetilde{c}_\gamma\right) \\
			c_{\gamma\gamma} &= -16 s_w^2 \bar{c}_\gamma &\quad&;\quad& \widetilde{c}_{\gamma\gamma} &= -16 s_w^2 \widetilde{c}_\gamma \\
			c_{Z\gamma} &= 2 \frac{s_w}{c_w} \left(\bar{c}_{HW} - \bar{c}_{HB} + 8 s_w^2 \bar{c}_\gamma\right) &\quad&;\quad& \widetilde{c}_{Z\gamma} &= 2 \frac{s_w}{c_w} \left(\widetilde{c}_{HW} - \widetilde{c}_{HB} + 8 s_w^2 \widetilde{c}_\gamma\right) \\
			c_{gg} &= 16 \frac{g_S^2}{g^2} \bar{c}_g &\quad&;\quad& \widetilde{c}_{gg} &= 16 \frac{g_S^2}{g^2} \widetilde{c}_g
		\end{alignedat}\end{equation}
		and:
		\begin{equation}\begin{aligned}
		\label{eq:kappaDico}
			\kappa_{Z\gamma} &= -2 \frac{s_w}{c_w} \left(\bar{c}_{HW} + \bar{c}_W - \bar{c}_{HB} - \bar{c}_B\right) \; , \\
			\kappa_{ZZ}      &= -2 \left(\bar{c}_{HW} + \bar{c}_W + \frac{s_w^2}{c_w^2} \bar{c}_{HB} + \frac{s_w^2}{c_w^2} \bar{c}_B\right) \; , \\
			\kappa_{WW}      &= -2 \left(\bar{c}_{HW} + \bar{c}_W\right) \; .
		\end{aligned}\end{equation}
		The SM Lagrangian is recovered when $c_W = c_Z = c_f = 1$ and all the $c_{ij} = 0$, $\widetilde{c}_{ij} = 0$ and $\kappa_{ij} = 0$. Due to the fact that we are working in the linear-realized EFT with a $SU(2)$ Higgs doublet, some of the parameters of~\ref{eq:L_1_eff} are related together; this translates into the following identities:
		\begin{align}
			\label{eq:cWW_with_cZZ_cZGam_cGamGam}
			c_{WW}             &= c_w^2 c_{ZZ}             + 2 c_w s_w c_{Z\gamma}             + s_w^2 c_{\gamma\gamma}             \; , \\
			\label{eq:cWWt_with_cZZt_cZGamT_cGamGamT}
			\widetilde{c}_{WW} &= c_w^2 \widetilde{c}_{ZZ} + 2 c_w s_w \widetilde{c}_{Z\gamma} + s_w^2 \widetilde{c}_{\gamma\gamma} \; , \\
			\label{eq:kWW_with_kZZ_kZGam}
			\kappa_{WW}        &= c_w^2 \kappa_{ZZ} + c_w s_w \kappa_{Z\gamma} \; ,
		\end{align}
		which happen to be the same as the ones obtained when imposing custodial symmetry by hand on a more general Higgs EFT where the hypotheses of linear realization of EWSB and the Higgs field representation are relaxed~\cite{Gonzalez-Alonso:2014eva}.

	\section{Additional Relations from Electroweak Corrections}
	\label{sec:ConstrParamsEWCorr}
		\par Higgs rates data publicly available from the ATLAS and CMS collaborations have up to now a limited power of discrimination between different tensor structures of the Higgs couplings to vector bosons. They cannot be used alone to put strong constraints on the effective Higgs couplings, therefore we need additional observables. A solution is to use existing constraints from EW precision observables and, in particular, we require that all the power divergences introduced in the oblique corrections by the existence of dimension-6 operators, should vanish.

		\par In many BSM models, new physics is due to heavy particles which intrinsic energy scale $M_{NP}$ is much larger than the EW scale $M_{EW}$ (for example, of the order of $M_\text{GUT}$ or $M_\text{Planck}$), so that no new possible EW-like gauge bosons can exist at $M_{EW}$ (apart from $\gamma$, $Z^0$ and $W^{\pm}$) and $SU(2)_L \times U(1)_Y$ still remains the EW gauge group~\cite{Peskin:1991sw,Barbieri:2004qk,Falkowski:2013dza}. The couplings of new physics particles to light fermions are required to be much smaller than those to gauge bosons. When those criteria are met, the dominant corrections due to the presence of new physics are corrections to the propagation of gauge bosons exchanged in 2-fermion scattering processes. These are the so-called electroweak oblique corrections conveniently parametrized by the Peskin-Takeuchi $S$, $T$, $U$ parameters~\cite{Peskin:1991sw}.

		\par Under the previous assumptions the gauge-boson two-point functions can be expanded around zero momentum in powers of $p^2$:

		\begin{equation}
			\Pi_{\mu\nu}(p^2) = g_{\mu\nu} \left( \Pi_{V_1 V_2}(p^2) = \Pi_{V_1 V_2}^{(0)}(0) + p^2 \Pi_{V_1 V_2}^{(2)}(0) + (p^2)^2 \Pi_{V_1 V_2}^{(4)}(0) + \cdots \right) + p_\mu p_\nu \left( \cdots \right)
		\end{equation}
		where the $V_i = 1,2,3,B$ label the $SU(2)_L \times U(1)_Y$ gauge bosons $W_{1,2,3}$ and $B$, and $\Pi_{V_1 V_2}^{(2k)}(0) \equiv \frac{1}{k!} \left(\frac{\partial}{\partial p^2}\right)^k \Pi_{V_1 V_2}\,(0)$. When denoting $\delta{\Pi}$ the shift of the corresponding two-point function from the SM value, the Peskin-Takeuchi parameters can be expressed (in terms of the $W_{1,2,3}$ and $B$, or $W^\pm$, $Z^0$ and $\gamma$ bosons) as:
		\begin{alignat}{3}
			\alpha S &= -4 s_w c_w \delta{\Pi_{3B}^{(2)}} &&= 4 s_w^2 c_w^2 \left( \delta{\Pi_{ZZ}^{(2)}} - \delta{\Pi_{\gamma\gamma}^{(2)}} - \frac{c_w^2 - s_w^2}{s_w c_w} \delta{\Pi_{Z\gamma}^{(2)}} \right) \; , \\
			\alpha T &= \frac{\delta{\Pi_{11}^{(0)}} - \delta{\Pi_{33}^{(0)}}}{m_W^2} &&= \frac{\delta{\Pi_{WW}^{(0)}}}{m_W^2} - \frac{c_w^2 \delta{\Pi_{ZZ}^{(0)}}}{m_W^2} = \frac{\delta{\Pi_{WW}^{(0)}}}{m_W^2} - \frac{\delta{\Pi_{ZZ}^{(0)}}}{m_Z^2} \; , \\
			\alpha U &= 4 s_w^2 \left( \delta{\Pi_{11}^{(2)}} - \delta{\Pi_{33}^{(2)}} \right) &&= 4 s_w^2 \left( \delta{\Pi_{WW}^{(2)}} - c_w^2 \delta{\Pi_{ZZ}^{(2)}} - s_w^2 \delta{\Pi_{\gamma\gamma}^{(2)}} - 2 c_w s_w \delta{\Pi_{Z\gamma}^{(2)}} \right) \; .
		\end{alignat}

\subsection{Tree-level constraints}
		\par Using $\mathcal{L}_0$ (Eq.~\ref{eq:L_0_eff}), the Lagrangian part that does not depend on the physical Higgs field $h$, we find:
		\begin{gather}
			\label{eq:TreeLevelSTU}
			\alpha S = 2 s_w^2 \left(\bar{c}_B + \bar{c}_W\right) \, , \quad \alpha T = \bar{c}_T \, , \quad \alpha U = 0 \, .
		\end{gather}

		At fixed $U = 0$, the up-to-date experimental limits~\cite{Baak:2014ora} for the $S$ and $T$ parameters, determined from a fit for a reference Standard Model with $m_{t,ref}=173$~GeV and $M_{H,ref}=125$~GeV, are:
		\begin{gather}
			S = \SI{0.06(9)}{} \, , \quad T = \SI{0.10(7)}{} \, ,
		\end{gather}
		corresponding to the following limits on the values of the combinations of the effective parameters:
		\begin{gather}
			\left|\bar{c}_B + \bar{c}_W\right| \leq \SI{1.0(15)e-3}{} \, , \quad 
			            \left|\bar{c}_T\right| \leq \SI{7.8(55)e-4}{} \, .       
		\end{gather}
		This motivates setting $\bar{c}_T$ and the combination $\bar{c}_B + \bar{c}_W$ to zero. In the next subsection we will see that it is indeed possible to be done.

\subsection{One-loop relations}
		\par In this subsection we evaluate at one-loop the self-energies of the form $V_1$--($V$/$H$)--$V_2$, namely: $Z$--($Z$/$H$)--$Z$ and $Z$--($\gamma$/$H$)--$Z$, $W$--($W$/$H$)--$W$, $\gamma$--($\gamma$/$H$)--$\gamma$ and $\gamma$--($Z$/$H$)--$\gamma$, and the $Z$/$\gamma$ mixing $Z$--($Z$/$H$)--$\gamma$ and $Z$--($\gamma$/$H$)--$\gamma$. To remain consistent with our usage of dimension-6 EFT we must consider only Higgs couplings corrections at linear order and not above; and for the EFT framework to remain valid, it is implicitely assumed that the cut-off scale $\Lambda$ of the EW loop is larger than $M_{EW}$ yet smaller than the new-physics scale $\Lambda_{NP}$, so that the effective $h V_1 V_2$ vertices can be still considered as "true" vertices (they are not resolved). We obtain the following power-divergent contributions to the $S$, $T$, $U$ parameters (see Appendices~\ref{subsec:1loop} and~\ref{subsec:STUCorr} for the details of the computations):

		\begin{adjustwidth}{-1cm}{}\resizemath{\small}
		\vspace{-0.5cm}
		\begin{align}
			\label{eq:aS}
			\alpha S = 
				&\; 2 s_w^2 c_w^2 \frac{\Lambda^2}{16 \pi^2 v^2} \left[
				\begin{aligned}
					&c_{ZZ}^2 - c_{\gamma\gamma}^2 - \widetilde{c}_{ZZ}^2 + \widetilde{c}_{\gamma\gamma}^2 + 2 c_Z \kappa_{ZZ} + 3 \left( c_{ZZ} \kappa_{ZZ} + c_{Z\gamma} \kappa_{Z\gamma} \right) \\
					&- \frac{c_w^2 - s_w^2}{s_w c_w} \left(c_Z \kappa_{Z\gamma} + c_{ZZ} c_{Z\gamma} + c_{Z\gamma} c_{\gamma\gamma} - \widetilde{c}_{ZZ} \widetilde{c}_{Z\gamma} - \widetilde{c}_{Z\gamma} \widetilde{c}_{\gamma\gamma}\right) \\
					&- \frac{c_w^2 - s_w^2}{s_w c_w} \frac{3}{2} \left(c_{Z\gamma} \kappa_{ZZ} + c_{\gamma\gamma} \kappa_{Z\gamma} + \kappa_{ZZ} \kappa_{Z\gamma}\right) + \frac{13 \kappa_{ZZ}^2 + 4 \kappa_{Z\gamma}^2}{3}
				\end{aligned}
				\right]
				+ \mathcal{O}\left(\ln{\widetilde{\Lambda}^2}\right) \, ,
			\\
			\label{eq:aT}
			\alpha T =
				&\; \frac{3}{8} \frac{\Lambda^4}{16 \pi^2 v^2} \left[\frac{\kappa_{ZZ}^2 + \kappa_{Z\gamma}^2}{m_Z^2}-\frac{\kappa_{WW}^2}{m_W^2}\right] + \frac{\Lambda^2}{16 \pi^2 v^2} \left[
				\begin{aligned}
					& c_Z^2 - c_W^2 + 3 c_Z \kappa_{ZZ} - 3 c_W \kappa_{WW} - 3 \frac{\kappa_{ZZ}^2 - \kappa_{WW}^2}{4} \\
					&- \frac{3 m_H^2}{4} \left(\frac{\kappa_{ZZ}^2 + \kappa_{Z\gamma}^2}{m_Z^2}-\frac{\kappa_{WW}^2}{m_W^2}\right)
				\end{aligned}
				\right]
				+ \mathcal{O}\left(\ln{\widetilde{\Lambda}^2}\right) \, ,
			\\
			\begin{split}
			\label{eq:aU}
			\alpha U =
				&\; 2 s_w^2 \frac{\Lambda^2}{16 \pi^2 v^2} \left[
				\begin{aligned}
					& c_{WW}^2 - \widetilde{c}_{WW}^2 + 3 c_{WW} \kappa_{WW} - 3 c_w^2 \left(c_{Z\gamma} \kappa_{Z\gamma} + c_{ZZ} \kappa_{ZZ}\right) - 3 c_w s_w \left(c_{\gamma\gamma} \kappa_{Z\gamma} + \kappa_{ZZ} \kappa_{Z\gamma} + c_{Z\gamma} \kappa_{ZZ}\right) \\
					&+ 2 \left(c_W \kappa_{WW} - c_w^2 c_Z \kappa_{ZZ} - c_w s_w c_Z \kappa_{Z\gamma}\right) + 13 \frac{\kappa_{WW}^2 - c_w^2 \kappa_{ZZ}^2}{3} - \frac{4 c_w^2 \kappa_{Z\gamma}^2}{3} \\
					&- \left(c_w c_{ZZ} + s_w c_{Z\gamma}\right)^2 - \left(c_w c_{Z\gamma} + s_w c_{\gamma\gamma}\right)^2 + \left(c_w \widetilde{c}_{ZZ} + s_w \widetilde{c}_{Z\gamma}\right)^2 + \left(c_w \widetilde{c}_{Z\gamma} + s_w \widetilde{c}_{\gamma\gamma}\right)^2
				\end{aligned}
				\right] \\
				&+ \mathcal{O}\left(\ln{\widetilde{\Lambda}^2}\right) \, .
			\end{split}
		\end{align}
		\endresizemath\end{adjustwidth}

		\par Given the current constraints on the oblique parameters~\cite{Baak:2014ora}, we require that loop-induced power-divergent corrections coming from the effective couplings cancel. We furthermore make a second hypothesis, namely that there are no fine-tuned cancellations between operators of different types, i.e. between the CP-even (generating the $c_V$ and $c_{VV}$ couplings) and CP-odd operators (generating the $\widetilde{c}_{VV}$ couplings) and the ones that generate the $\kappa_i$ couplings.

		\par The cancellation of the quartic divergence in the $T$ parameter requires\footnote{Using: $c_w^2 = \frac{m_W^2}{m_Z^2}$.} that $\kappa_{WW}^2 = c_w^2 \left(\kappa_{ZZ}^2 + \kappa_{Z\gamma}^2\right)$. However raising the constraint~\ref{eq:kWW_with_kZZ_kZGam} to the square gives: $\kappa_{WW}^2 = c_w^2 \left(c_w \kappa_{ZZ} + s_w \kappa_{Z\gamma}\right)^2$, so that we obtain: $\kappa_{ZZ}^2 + \kappa_{Z\gamma}^2 = \left(c_w \kappa_{ZZ} + s_w \kappa_{Z\gamma}\right)^2$. After expansion of the right-hand side and rewrite of this equation, we obtain: $\left(s_w \kappa_{ZZ} - c_w \kappa_{Z\gamma}\right)^2 = 0$, so that $\kappa_{Z\gamma} = \frac{s_w}{c_w} \kappa_{ZZ}$. Reinjecting this relation in~\ref{eq:kWW_with_kZZ_kZGam} gives: $\kappa_{WW} = \kappa_{ZZ}$. Therefore each of the power-divergent part of the $S$, $T$, $U$ parameters can be rewritten under the form:
		\begin{equation}
			S,T,U = \frac{\Lambda^2}{16 \pi^2 v^2} \mathcal{P}(c_i, \widetilde{c}_i, \kappa_{ZZ}) + \mathcal{O}\left(\ln{\widetilde{\Lambda}^2}\right)
		\end{equation}
		where $\mathcal{P}$ is a polynomial of degree 2. The $T$ parameter can be rewritten as:
		\begin{equation}
			\alpha T = \frac{\Lambda^2}{16 \pi^2 v^2} \left[c_Z^2 - c_W^2 + 3 \kappa_{ZZ} (c_Z - c_W)\right] + \mathcal{O}\left(\ln{\widetilde{\Lambda}^2}\right)
		\end{equation}
		and removing the quadratic divergence from it can be done if $c_Z = c_W \equiv c_V$ (first custodial relation).

		\par The $U$ parameter can be reexpressed as:
		\resizemath{\small}
		\begin{equation}\begin{split}
			\label{eq:aU}
			\alpha U =
				&\; 2 s_w^2 \frac{\Lambda^2}{16 \pi^2 v^2} \left[
				\begin{aligned}
					& c_{WW}^2 - \widetilde{c}_{WW}^2 - \left(c_w c_{ZZ} + s_w c_{Z\gamma}\right)^2 - \left(c_w c_{Z\gamma} + s_w c_{\gamma\gamma}\right)^2 \\
					&+ \left(c_w \widetilde{c}_{ZZ} + s_w \widetilde{c}_{Z\gamma}\right)^2 + \left(c_w \widetilde{c}_{Z\gamma} + s_w \widetilde{c}_{\gamma\gamma}\right)^2 \\
					& + 3 \kappa_{ZZ} \left(c_{WW} - c_w^2 c_{ZZ} - s_w^2 c_{\gamma\gamma} - 2 s_w c_w c_{Z\gamma}\right)
				\end{aligned}
				\right]
				+ \mathcal{O}\left(\ln{\widetilde{\Lambda}^2}\right) \, .
		\end{split}\end{equation}
		\endresizemath

		The constraint~\ref{eq:cWW_with_cZZ_cZGam_cGamGam} automatically cancels the last line of the previous equation, so that removing all the remaining power divergences in $U$ without fine-tuning between the CP-even and CP-odd parts implies:
		\begin{equation}\begin{aligned}
			\label{eq:cWWSq_and_cWWtSq}
			c_{WW}^2 &= \left(c_w c_{ZZ} + s_w c_{Z\gamma}\right)^2 + \left(c_w c_{Z\gamma} + s_w c_{\gamma\gamma}\right)^2 \, , \\
			\widetilde{c}_{WW}^2 &= \left(c_w \widetilde{c}_{ZZ} + s_w \widetilde{c}_{Z\gamma}\right)^2 + \left(c_w \widetilde{c}_{Z\gamma} + s_w \widetilde{c}_{\gamma\gamma}\right)^2 \, .
		\end{aligned}\end{equation}
		By squaring the constraints~\ref{eq:cWW_with_cZZ_cZGam_cGamGam} and~\ref{eq:cWWt_with_cZZt_cZGamT_cGamGamT} and equating them with the previous equations, we obtain: $c_{Z\gamma}^2 \left(1 - 4 c_w^2 s_w^2\right) - 2 c_{Z\gamma} \frac{c_w^2 - s_w^2}{s_w c_w} \left(c_{ZZ} - c_{\gamma\gamma}\right) + \left(c_{ZZ} - c_{\gamma\gamma}\right)^2 = 0$ (and similar for $\widetilde{c}_i$), and since:~$\left(1 - 4 c_w^2 s_w^2\right) = \left(\frac{c_w^2 - s_w^2}{s_w c_w}\right)^2$, this translates to:
		\begin{equation}\begin{alignedat}{3}
		\label{eq:cZGam_and_cZGamT}
			c_{Z\gamma} &= \frac{s_w c_w}{c_w^2 - s_w^2} \left( c_{ZZ} - c_{\gamma\gamma} \right) &\quad&\rightarrow\quad& c_{ZZ} &= c_{\gamma\gamma} + \frac{c_w^2 - s_w^2}{s_w c_w} c_{Z\gamma} \, , \\
			\widetilde{c}_{Z\gamma} &= \frac{s_w c_w}{c_w^2 - s_w^2} \left( \widetilde{c}_{ZZ} - \widetilde{c}_{\gamma\gamma} \right) &\quad&\rightarrow\quad& \widetilde{c}_{ZZ} &= \widetilde{c}_{\gamma\gamma} + \frac{c_w^2 - s_w^2}{s_w c_w} \widetilde{c}_{Z\gamma} \, .
		\end{alignedat}\end{equation}
		Hence, Eqs.~\ref{eq:cWWSq_and_cWWtSq} are verified, and we removed all the quadratic divergences in $U$. Reinjecting these relations into~\ref{eq:cWW_with_cZZ_cZGam_cGamGam} and~\ref{eq:cWWt_with_cZZt_cZGamT_cGamGamT} give:
		\begin{equation}
			c_{WW} = c_{\gamma\gamma} + \frac{c_w}{s_w} c_{Z\gamma} \quad\text{and}\quad \widetilde{c}_{WW} = \widetilde{c}_{\gamma\gamma} + \frac{c_w}{s_w} \widetilde{c}_{Z\gamma} \, .
		\end{equation}

		\par The $S$ parameter can be reexpressed as:
		\resizemath{\small}
		\begin{equation}\begin{split}
			\label{eq:aS}
			\alpha S =
				&\; 2 s_w^2 c_w^2 \frac{\Lambda^2}{16 \pi^2 v^2} \left[
				\begin{aligned}
					& c_{ZZ}^2 - c_{\gamma\gamma}^2 - \widetilde{c}_{ZZ}^2 + \widetilde{c}_{\gamma\gamma}^2 - \frac{c_w^2 - s_w^2}{s_w c_w} \left( c_{ZZ} c_{Z\gamma} + c_{Z\gamma} c_{\gamma\gamma} - \widetilde{c}_{ZZ} \widetilde{c}_{Z\gamma} - \widetilde{c}_{Z\gamma} \widetilde{c}_{\gamma\gamma} \right) \\
					&+ c_V \kappa_{ZZ} \left(1 + \frac{s_w^2}{c_w^2}\right) + 3 \kappa_{ZZ} \left(c_{ZZ} + \frac{c_{Z\gamma}}{2} \left(3 \frac{s_w}{c_w} - \frac{c_w}{s_w}\right)\right) \\
					&- \frac{3}{2} c_{\gamma\gamma} \kappa_{ZZ} \left(1 - \frac{s_w^2}{c_w^2}\right) + \frac{17}{6} \kappa_{ZZ}^2 \left(1 + \frac{s_w^2}{c_w^2}\right)
				\end{aligned}
				\right]
				+ \mathcal{O}\left(\ln{\widetilde{\Lambda}^2}\right) \, .
		\end{split}\end{equation}
		\endresizemath
		All the power divergences in $S$ disappear without fine-tuning between the CP-even and CP-odd parts if the following relations are satisfied:
		\begin{equation}\begin{aligned}
			c_{ZZ}^2 &= c_{\gamma\gamma}^2 + \frac{c_w^2 - s_w^2}{s_w c_w} \left( c_{ZZ} c_{Z\gamma} + c_{Z\gamma} c_{\gamma\gamma} \right) \, , \\
			\widetilde{c}_{ZZ}^2 &= \widetilde{c}_{\gamma\gamma}^2 + \frac{c_w^2 - s_w^2}{s_w c_w} \left( \widetilde{c}_{ZZ} \widetilde{c}_{Z\gamma} + \widetilde{c}_{Z\gamma} \widetilde{c}_{\gamma\gamma} \right) \, .
		\end{aligned}\end{equation}
		However these equations are precisely~\ref{eq:cZGam_and_cZGamT}, multiplied by $c_{ZZ} + c_{\gamma\gamma}$ or $\widetilde{c}_{ZZ} + \widetilde{c}_{\gamma\gamma}$ respectively. Therefore the first line in $\alpha S$ cancels and only a quadratic divergence proportional to $\kappa_{ZZ}$ remains. The only way to remove it is to put $\kappa_{ZZ} = 0$. Another alternative would be to keep only the $c_i$ and $\widetilde{c}_i$ terms by taking\footnote{This can be motivated by computing the $W$-parameter, introduced by Barbieri, Pomarol, Rattazzi and Strumia~\cite{Barbieri:2004qk}. Its quadratic divergence is proportional to $\kappa_{WW}^2 \equiv \kappa_{ZZ}^2$. Requiring its vanishing implies $\kappa_i = 0$.} $\kappa_{ZZ} = 0$ right from the beginning, so that all of the $\kappa_i = 0$. Doing so we retrieve the same constraints as before.

		\par Only pure logarithmic divergences therefore remain in the Peskin-Takeuchi parameters (see Appendix~\ref{subsec:STUCorr} for details) by using these relations, summarized here:
		\begin{align}
			\label{eq:ki}    \kappa_i &= 0 \, , \\
			\label{eq:cV}         c_Z &= c_W \equiv c_V \, , \\
			\label{eq:cWWcWWt} c_{WW} &= c_{\gamma\gamma} + \frac{c_w}{s_w} c_{Z\gamma} \quad \text{and similar for} \; \widetilde{c}_{WW} \, , \\
			\label{eq:cZZcZZt} c_{ZZ} &= c_{\gamma\gamma} + \frac{c_w^2 - s_w^2}{s_w c_w} c_{Z\gamma} = c_{WW} - \frac{s_w}{c_w} c_{Z\gamma} \quad \text{and similar for} \; \widetilde{c}_{ZZ} \, .
		\end{align}
		Those relations can be seen as extended custodial relations that link the parameters of the effective Lagrangian~\ref{eq:L_1_eff}, and they are equivalent to the following constraints on the SILH Lagrangian parameters:
\vspace{-0.5cm}
		\begin{alignat}{3}
			\label{eq:C1}
			(\ref{eq:ki}) &\quad&\rightarrow\quad&& \bar{c}_{HB} + \bar{c}_B &= 0 = \bar{c}_{HW} + \bar{c}_W \, , \\
			\label{eq:C2}
			(\ref{eq:cV}) &\quad&\rightarrow\quad&& \bar{c}_T &= 0 \, , \\
			\label{eq:C3}
			(\ref{eq:cWWcWWt}) \;\text{or}\; (\ref{eq:cZZcZZt}) &\quad&\rightarrow\quad&& \bar{c}_{HW} + \bar{c}_{HB} &= 0 = \widetilde{c}_{HW} + \widetilde{c}_{HB} \, ,
		\end{alignat}
		leading in particular to:
		\begin{equation}
			\bar{c}_{HW} = - \bar{c}_W = - \bar{c}_{HB} = \bar{c}_B \quad \text{or:} \quad \bar{c}_B + \bar{c}_W = 0 \, .
		\end{equation}
		Together with~\ref{eq:C2} and the tree-level relation~\ref{eq:TreeLevelSTU}, this means that the theory only generates oblique corrections starting at one-loop but not at tree-level, and our guess from the previous subsection is justified.

		\par Summarizing, after imposing electroweak constraints, the effective Higgs Lagrangian~\ref{eq:L_1_eff} depends on 7 independent parameters in the CP-even sector:
		\begin{equation}
			\label{eq:CPEvenParameters}
			c_V, \quad c_u, \quad c_d, \quad c_l, \quad c_{\gamma\gamma}, \quad c_{Z\gamma}, \quad c_{gg}
		\end{equation}
		and 6 independent parameters in the CP-odd sector:
		\begin{equation}
			\label{eq:CPOddParameters}
			\widetilde{c}_u, \quad \widetilde{c}_d, \quad \widetilde{c}_l, \quad \widetilde{c}_{\gamma\gamma}, \quad \widetilde{c}_{Z\gamma}, \quad \widetilde{c}_{gg}
		\end{equation}
		and the tree-level SM Higgs Lagrangian is retrieved when $c_V = c_{f=u,d,l} = 1$, $c_{gg} = c_{\gamma\gamma} = c_{Z\gamma} = 0$ and all the $\widetilde{c}_{ij} = 0$, while the $c_{ij}$ are generated only at loop-level.

\section{Comparison of the theory with experimental data}
	\par The LHC experiments usually provide measurements of the relative Higgs decay rates (signal strengths) in various channels, defined as: $\hat{\mu}^{YH}_{XX} = \frac{\sigma_{YH}}{\sigma_{YH}^{SM}} \frac{{\rm Br}(h \to XX)}{{\rm Br}(h \to XX)_{SM}}$. The relative branching fraction reads: $\frac{{\rm Br}(h \to XX)}{{\rm Br}(h \to XX)_{SM}} = \frac{\Gamma_{XX}}{\Gamma_{XX,SM}} \frac{\Gamma_{tot,SM}}{\Gamma_{tot}}$, where $\Gamma_{tot}$ is the sum of all the partial widths. The efficiencies of analysis cuts applied on final states are absorbed into the definition of the cross-sections. An effect of the effective operators is that both Higgs decay rates and production cross-sections in those channels are shifted from their SM values. Hence the parameters of the effective Lagrangian can be constrained by comparing the theoretical rates from the SM with the measured ones. In the following we summarize how they depend on the parameters of the effective Lagrangian. We do not consider any contributions to the Higgs width other than Higgs decays into SM particles.

	\par We use the following values for the SM constants (from PDG~2015~\cite{Agashe:2014kda}):
	\begin{equation}\begin{gathered}
		G_F = \SI{1.16637e-5}{GeV^{-2}}, \quad \alpha_{EW}^{-1}(m_Z) = \num{128.462}, \quad \alpha_S(m_Z;m_h) =\hspace{-1.5mm}\footnotemark\; (\num{0.1184};\num{0.1122}), \\
		\left( g_L(m_Z;m_h) = (\num{0.657448};\num{0.643133}), \quad g_Y(m_Z;m_h) = (\num{0.341167};\num{0.357943}) \right), \\
		m_Z = \SI{91.1875}{GeV}, \quad (m_W = \SI{80.385}{GeV}), \quad m_h = \SI{125.09}{GeV}, \\
		m_c(m_c;m_h) = (\num{1.29};\num{0.61593})\si{GeV}, \quad m_b(m_b;m_h) = (\num{4.6};\num{2.7636})\si{GeV}, \quad m_t(m_h;m_t) = (\num{169.16};\num{173.2})\si{GeV}, \\
		\Gamma_t = \SI{1.5083}{GeV}, \quad \Gamma_Z = \SI{2.4952}{GeV}, \quad \Gamma_W = \SI{2.085}{GeV}, \quad \Gamma_h = \SI{4.154}{MeV}.
	\end{gathered}\end{equation}
	\footnotetext{World's average of $\alpha_S$ in 2012, see~\cite{Bethke:2012jm}.}

	\subsection{Relative decay widths}
	\label{subsec:RelDecWidth}
		\par The on-shell decay rate of the Higgs boson into a (charged) fermion $f$ and its antifermion, at tree-level, can be written as:
		\begin{equation}\begin{split}
			\left(\frac{\Gamma}{\Gamma_{SM}}\right)_{h \to f \overline{f}} &= \left|c_f\right|^2 + \left|\widetilde{c}_f\right|^2 \frac{m_h^2}{m_h^2 - 4 m_f^2} \\
			                                                               &\simeq \left|c_f\right|^2 + \left|\widetilde{c}_f\right|^2 \quad \text{if $f \neq$ top quark.}
		\end{split}\end{equation}

		\par The decay rate of the Higgs into two vector bosons $V_1$ and $V_2$, where $(V_1, V_2) = (g, g), (\gamma ,\gamma) \;\text{or}\; (Z, \gamma)$, is generated starting one-loop level in the SM and receives a tree-level contribution from the corresponding effective coupling. It can be written as:
		\begin{equation}
			\left(\frac{\Gamma}{\Gamma_{SM}}\right)_{h \to V_1 V_2} \simeq \frac{\left|\widehat{c_{V_1 V_2}}\right|^2 + \left|\widehat{\widetilde{c}_{V_1 V_2}}\right|^2}{\left|\widehat{c_{V_1 V_2 , SM}}\right|^2} \, .
		\end{equation}
		The hatted quantities are effective CP-even and CP-odd "bookkeeping" Higgs couplings to $V_1$ and $V_2$ which include two types of contributions that enter the decay amplitude at the same order in the effective theory: the tree-level contributions proportional to the effective couplings $c_{ij}$ of the NLO Lagrangian, and the one-loop quantum corrections, proportional to the effective couplings $c_i$, $\widetilde{c}_i$ of the LO Lagrangian (which also include SM corrections). The hatted quantity noted "SM" is the value of the corresponding hatted effective coupling when using SM values for the $c_i$ couplings. For our level of precision, we can safely stop the expansion of the relative decay widths (and the relative cross-sections, see after) at NLO, because of the following reason: for $h \to \gamma\gamma$ and $h \to Z\gamma$, the NLO corrections arise in QED only and are of order $0.2\%$, and for $h \to gg$, even if QCD N$^{2+}$LO corrections are sizeable, they almost cancel in the relative signal strengths and leave only a $\approx 2\%$ correction~\cite{Gori:2013mia}.

		\par We have implemented this effective model\footnote{During the completion of this work, \cite{Alloul:2013naa} appeared where the complete SILH Lagrangian was implemented as a full \Verb+FeynRules+ model. Our goal was simpler and we only implemented the needed operators after expansion in the mass basis.} with \Verb+FeynRules 2.1+~\cite{Christensen:2008py,Alloul:2013bka} and we used \Verb+FeynArts 3.9+~\cite{Kublbeck:1990xc,Hahn:2000kx} and \Verb+FormCalc 8.4+~\cite{Nejad:2013ina,Gross:2014ola} to generate and numerically evaluate the relative decay widths. We obtain:

		\begin{itemize}
			\item[\textbullet] \uline{$V_1 = V_2 = g$}:
				\begin{equation}\begin{aligned}
				\label{eq:RelWidthGGH}
					\widehat{c_{gg}}             &\simeq c_{gg} + 10^{-2} 1.226 c_t - 10^{-4}(3.868-4.175\,\imath)c_b \, , \\
					\widehat{\widetilde{c}_{gg}} &\simeq \widetilde{c}_{gg} - 10^{-2} 1.868 \widetilde{c}_t + 10^{-4}(4.225-4.183\,\imath)\widetilde{c}_b \, , \\
					\left|\widehat{c_{gg, SM}}\right| &\simeq 0.0119 \, , 
				\end{aligned}\end{equation}
				which gives also the relative production cross-section via gluon-fusion $\frac{\sigma_{ggh}}{\sigma_{ggh,SM}}$.

			\item[\textbullet] \uline{$V_1 = V_2 = \gamma$}:
				\begin{equation}\begin{aligned}
					\widehat{c_{\gamma\gamma}}             &\simeq c_{\gamma\gamma} + 10^{-2}(1.032 c_V - 0.227 c_t) + 10^{-5}(1.790-1.932\,\imath)c_b + 10^{-5}(2.954-2.684\,\imath)c_\tau \, , \\
					\widehat{\widetilde{c}_{\gamma\gamma}} &\simeq \widetilde{c}_{\gamma\gamma} + 10^{-3} 3.458 \widetilde{c}_t - 10^{-5}(1.955-1.935\,\imath)\widetilde{c}_b - 10^{-5}(3.157-2.686\,\imath)\widetilde{c}_\tau \, , \\
					\left|\widehat{c_{\gamma\gamma, SM}}\right| &\simeq 0.0081 \, . 
				\end{aligned}\end{equation}
				\vspace{-1cm}

			\item[\textbullet] \uline{$V_1 = Z, V_2 = \gamma$}:
				\begin{equation}\begin{aligned}
					\widehat{c_{Z\gamma}}             &\simeq c_{Z\gamma} + 10^{-2}(1.472 c_V - 0.0763 c_t) + 10^{-6}(8.452-4.136\,\imath)c_b + 10^{-7}(3.144-1.350\,\imath)c_\tau \, , \\
					\widehat{\widetilde{c}_{Z\gamma}} &\simeq \widetilde{c}_{Z\gamma} + 10^{-3} 1.158 \widetilde{c}_t - 10^{-6}(9.641-4.140\,\imath)\widetilde{c}_b - 10^{-7}(3.530-1.351\,\imath)\widetilde{c}_\tau \, , \\
					\left|\widehat{c_{Z\gamma, SM}}\right| &\simeq 0.0140 \, .
				\end{aligned}\end{equation}
				\vspace{-1cm}
		\end{itemize}

		\par At tree-level, the Higgs boson decays also into $ZZ^*$ or $WW^*$, subsequently decaying into leptons. The corresponding relative decay widths were computed with \Verb+MadGraph 5+~\cite{Alwall:2011uj,Alwall:2014hca}, 
		using the output in the UFO~format~\cite{Degrande:2011ua} generated from our \Verb+FeynRules+ model. The following cut selection was chosen: $m_{\ell\ell} > \SI{4}{GeV}$ on the invariant mass of each pair of charged leptons produced by the decay of the $Z$ or from the $W$'s. The effects of one-loop corrections were implemented in \Verb+MadGraph+ computations by using the hatted effective couplings. Using the extended-custodial relations~\ref{eq:cV}, \ref{eq:cWWcWWt} and~\ref{eq:cZZcZZt} the expressions are written in terms of the CP-even variables $c_V$, $c_{\gamma\gamma}$, $c_{Z\gamma}$ and the corresponding CP-odd ones. After variation of the values of the $c_i$ and $\widetilde{c}_i$ couplings and performing a fit on a polynomial of the form given in Eq.~\ref{eq:RelXSecForm}, we obtain the following numerical estimations:
		\begin{align}
			\begin{split}
			\label{eq:RelWidthHZZ}
			\left(\frac{\Gamma}{\Gamma_{SM}}\right)_{ZZ^* \to 4\ell}     \simeq
				c_V^2 & + 0.0507 c_{\gamma\gamma}^2 + 0.0780 c_{Z\gamma}^2 + 0.4404 c_V c_{\gamma\gamma} + 0.5462 c_V c_{Z\gamma} + 0.1258 c_{\gamma\gamma} c_{Z\gamma} \\
					  & + 0.0116 \widetilde{c}_{\gamma\gamma}^2 + 0.0178 \widetilde{c}_{Z\gamma}^2 + 0.0288 \widetilde{c}_{\gamma\gamma} \widetilde{c}_{Z\gamma}
			\end{split} \, , \\
			\begin{split}
			\label{eq:RelWidthHWW}
			\left(\frac{\Gamma}{\Gamma_{SM}}\right)_{WW^* \to 2\ell2\nu} \simeq
				c_V^2 & + 0.0811 c_{\gamma\gamma}^2 + 0.2619 c_{Z\gamma}^2 + 0.5507 c_V c_{\gamma\gamma} + 0.9896 c_V c_{Z\gamma} + 0.2914 c_{\gamma\gamma} c_{Z\gamma} \\
					  & + 0.0207 \widetilde{c}_{\gamma\gamma}^2 + 0.0668 \widetilde{c}_{Z\gamma}^2 + 0.0743 \widetilde{c}_{\gamma\gamma} \widetilde{c}_{Z\gamma}
			\end{split} \, .
		\end{align}
		Changing this choice of cut to $m_{\ell\ell} > \SI{12}{GeV}$, only changes the numerical factors at the level of $\num{e-3}$.

	\subsection{Relative production cross-sections}
		\par For gluon-fusion production mode, the expression of the relative production cross-section $\frac{\sigma_{ggh}}{\sigma_{ggh,SM}}$ is taken to be the same as the one of the decay rate via gluon-fusion, Eq.~\ref{eq:RelWidthGGH}.

		\par In the case of the "Vector boson fusion" (VBF) or "Vector boson associated" (VH) production modes, we simulated the production of Higgs via $pp$ collisions at $\sqrt{s} = \SI{8}{TeV}$ with \Verb+MadGraph+ using \Verb+NNPDF23LO1+ PDF set~\cite{Ball:2012cx,Deans:2013mha}; then the same procedure as for the relative widths~\ref{eq:RelWidthHZZ} and~\ref{eq:RelWidthHWW} was used: after variation of the effective couplings the relative cross-sections was fitted with a polynomial of the form:
		\begin{equation}
		\label{eq:RelXSecForm}
			\left(\frac{\sigma}{\sigma_{SM}}\right) \simeq
				c_V^2 + \alpha_1 c_{\gamma\gamma}^2 + \alpha_2 c_{Z\gamma}^2 + \alpha_3 c_V c_{\gamma\gamma} + \alpha_4 c_V c_{Z\gamma} + \alpha_5 c_{\gamma\gamma} c_{Z\gamma} + \beta_1 \widetilde{c}_{\gamma\gamma}^2 + \beta_2 \widetilde{c}_{Z\gamma}^2 + \beta_3 \widetilde{c}_{\gamma\gamma} \widetilde{c}_{Z\gamma} \, .
		\end{equation}
		\par As said before, we absorb the cuts efficiencies into the cross-sections. This means that the expressions for the relative cross-sections are really $\frac{\sigma \cdot \epsilon}{\sigma_{SM} \cdot \epsilon_{SM}}$, where $\epsilon$ and $\epsilon_{SM}$ are the cuts efficiencies for (SM+BSM) and (SM) only. It is usually assumed that $\epsilon = \epsilon_{SM}$, which is true if BSM Higgs couplings are proportional to SM Higgs couplings, but generally it is not when BSM physics introduces new tensor structures in the couplings, and this is what happens in our case when taking into account the Higgs couplings to product of field-strength tensors, and the presence of CP-odd couplings~\cite{Banerjee:2013apa}. Therefore the coefficients $\alpha_i$ and $\beta_i$ of Eq.~\ref{eq:RelXSecForm} depend on the sets of cuts on the kinematics of the final-state jets chosen to perform the analysis/selection.

		\begin{itemize}

			\item[\textbullet] Vector boson fusion (VBF): $qq \to hqq$ with exchange of $W$ or $Z$ bosons. To compute the coefficients we choose a common cut setting for both ATLAS and CMS: $m_{jj} > \SI{250}{GeV}$, $\eta < 5$, $p_T > \SI{20}{GeV}$, which is relaxed enough to be able to cover both ATLAS and CMS choices. A cross-check was done to verify that the coefficients were consistent with the ones obtained by using dedicated ATLAS or CMS cuts; they are modified at most up to $20\%$ level.
			\begin{equation}
				\begin{split}
				\left(\frac{\sigma}{\sigma_{SM}}\right)_{VBF} \simeq
					c_V^2 & + 1.816 c_{\gamma\gamma}^2 + 3.796 c_{Z\gamma}^2 + 0.351 c_V c_{\gamma\gamma} + 0.623 c_V c_{Z\gamma} + 4.140 c_{\gamma\gamma} c_{Z\gamma} \\
						  & + 1.555 \widetilde{c}_{\gamma\gamma}^2 + 3.073 \widetilde{c}_{Z\gamma}^2 + 3.356 \widetilde{c}_{\gamma\gamma} \widetilde{c}_{Z\gamma}
				\end{split} \, .
			\end{equation}

			\item[\textbullet] Vector boson associated production (VH): $q\bar{q} \to hV$, where $V = W,Z$. By using an inclusive cut, we get:
			\begin{align}
				\begin{split}
				\left(\frac{\sigma}{\sigma_{SM}}\right)_{hW} \simeq
					c_V^2 & + 5.284 c_{\gamma\gamma}^2 + 17.058 c_{Z\gamma}^2 - 3.597 c_V c_{\gamma\gamma} - 6.463 c_V c_{Z\gamma} + 18.987 c_{\gamma\gamma} c_{Z\gamma} \\
						  & + 3.229 \widetilde{c}_{\gamma\gamma}^2 + 10.439 \widetilde{c}_{Z\gamma}^2 + 11.611 \widetilde{c}_{\gamma\gamma} \widetilde{c}_{Z\gamma}
				\end{split} \, , \\
				\begin{split}
				\left(\frac{\sigma}{\sigma_{SM}}\right)_{hZ} \simeq
					c_V^2 & + 4.197 c_{\gamma\gamma}^2 + 12.100 c_{Z\gamma}^2 - 3.324 c_V c_{\gamma\gamma} - 5.079 c_V c_{Z\gamma} + 12.897 c_{\gamma\gamma} c_{Z\gamma} \\
						  & + 2.374 \widetilde{c}_{\gamma\gamma}^2 + 6.937 \widetilde{c}_{Z\gamma}^2 + 7.317 \widetilde{c}_{\gamma\gamma} \widetilde{c}_{Z\gamma}
				\end{split} \, .
			\end{align}
			In particular for ZH production, we allow the presence of a $\gamma$ in s-channel, due to the existence of the effective $h-Z-\gamma$ vertex.

		\end{itemize}

\section{Experimental data}
	\par In our global fit we include the latest Run-I LHC results from the ATLAS and CMS experiments, summarized in Table~\ref{tab:exp}. In the case of the $Z\gamma$ decay channel where CMS provides only $95\%$~CL limits, we reconstruct its $\hat\mu$ assuming Gaussian errors. For some decay channels, the experiments provide the full 2-dimensional (2D) likelihood functions defined in the $\hat\mu_{ggH+ttH}$--$\hat\mu_{VBF+VH}$ plane; we use them because they encode the non-trivial correlations between the rates measured for the $ggH$/$ttH$ or VBF/VH production modes. In this case we provide for illustration purposes only a value of $\hat\mu$ obtained after a basic recombination of $\hat\mu_{ggH+ttH}$ and $\hat\mu_{VBF+VH}$. When only $95\%$~CL or $68\%$~CL contours of the 2D likelihoods are given for these channels, we reconstruct an approximate 2D likelihood function in the whole $\hat\mu_{ggH+ttH}$--$\hat\mu_{VBF+VH}$ plane by using a quadratic likelihood polynomial in those two variables, built such that its section corresponding to $95\%$~CL is an ellipsis that fits the contour.

	\par In addition to LHC data we use electroweak precision measurements from LEP, SLC and Tevatron, which are collected and can be found in Table~1 of Falkowski {\it et~al.}~\cite{Falkowski:2013dza}. Higgs loops introduce logarithmically divergent corrections, function of a cut-off scale, to the electroweak precision observables; they should therefore be evaluated with an explicit value of the cut-off: we assume $\Lambda_{NP} = 3$~TeV.

\vspace{0.5cm}

\begin{table}[h!]
\centering
\caption{
The LHC Higgs rates included in the fit.
The "2D" production holds for ggH+ttH and VBF+VH production modes, whose likelihood functions are defined in the plane $\mu_{\rm ggh+tth}$-$\mu_{\rm VBF+Vh}$. For the the diphoton channel (cats.) we use the five-dimensional likelihood function in the space spanned by $(\mu_{ggh}, \mu_{tth}, \mu_{\rm VBF}, \mu_{Wh},\mu_{Zh})$. For these two cases $\mu$ is quoted for illustration only. Correlations amongst different production classes in this table are ignored.}
\label{tab:exp}

\renewcommand*{\arraystretch}{1.2} 

\begin{tabular}{|c|c|c|c|c|c|}
\hline
Channel & $\mu_{\rm{ATLAS}}$ & $\mu_{\rm{CMS}}$ & $\mu_\text{Comb}$ & Production & Ref. \\
\hline
$\gamma \gamma$ & $1.17^{+0.28}_{-0.26}$ & $1.12^{+0.25}_{-0.22}$ & - & cats. & \cite{Aad:2014eha,Khachatryan:2014ira} \\
\hline
$Z \gamma$ & $2.7^{+4.5}_{-4.3}$ & $-0.2^{+4.9}_{-4.9}$ & - & total   &  \cite{Aad:2015gba,Chatrchyan:2013vaa} \\
\hline
$Z Z^*$ & $1.46^{+0.40}_{-0.34}$ & $1.00^{+0.29}_{-0.29}$ & $1.31^{+0.27}_{-0.14}$ & 2D & \cite{Aad:2014eva,Khachatryan:2014jba,LHCcomb} \\
\hline
$W W^*$ & $1.18^{+0.24}_{-0.21}$ & $0.83^{+0.21}_{-0.21}$ & $1.11^{+0.18}_{-0.17}$ & 2D & \cite{ATLAS:2014aga,Khachatryan:2014jba,LHCcomb} \\
\cline{2-6}
	& $2.1^{+1.9}_{-1.6}$ & - & - & Wh & \cite{Aad:2015ona} \\
\cline{2-6}
	& $5.1^{+4.3}_{-3.1}$ & - & - & Zh & \cite{Aad:2015ona} \\
\cline{2-6}
	& - & $0.80^{+1.09}_{-0.93}$ & - & Vh & \cite{Khachatryan:2014jba} \\
\hline
$\tau \tau$ & $1.44^{+0.42}_{-0.37}$ & $0.91^{+0.28}_{-0.28}$ & $1.12^{+0.25}_{-0.23}$ & 2D & \cite{Aad:2015vsa,Khachatryan:2014jba,LHCcomb} \\
\cline{2-6}
	& - & $0.87^{+1.00}_{-0.88}$ & - & Vh & \cite{Khachatryan:2014jba} \\
\hline
$b b$  & $1.11^{+0.65}_{-0.61}$ & - & - & Wh & \cite{Aad:2014xzb} \\
\cline{2-6}
	& $0.05^{+0.52}_{-0.49}$ & - & - & Zh & \cite{Aad:2014xzb} \\
\cline{2-6}
	& - & $0.89^{+0.47}_{-0.44}$ & - & Vh & \cite{Khachatryan:2014jba} \\
\cline{2-6}
	& - & $2.8^{+1.6}_{-1.4}$ & - & VBF & \cite{Khachatryan:2015bnx} \\
\cline{2-6}
	& $1.5^{+1.1}_{-1.1}$ & $1.2^{+1.6}_{-1.5}$ & - & tth & \cite{Aad:2015gra,Khachatryan:2015ila} \\
\hline
$\mu \mu$  & $-0.7^{+3.7}_{-3.7}$ & $0.8^{+3.5}_{-3.4}$ & - & total & \cite{Aad:2015gba,Khachatryan:2014aep} \\
\hline
multi-$\ell$  & $2.1^{+1.4}_{-1.2}$ & $3.8^{+1.4}_{-1.4}$ & - & tth & \cite{Aad:2015iha,Khachatryan:2014qaa} \\
\hline
\end{tabular}

\end{table}

\section{Global fit -- Discussion}
	\par We use the previous definitions for building the theoretical expressions of the signal strengths $\hat\mu^{th}$ for different channels, which depend on the effective couplings $c_i$ and $\widetilde{c}_i$. For the decay channels where we only know each signal strength $\hat\mu^{exp} \pm \delta\mu$ separately, we assume the errors to be Gaussian and uncorrelated and we define a 1-dimensional chi-squared function: $\chi_{1D}^2(\hat\mu^{th}, \hat\mu^{exp} \pm \delta\mu) = \left(\frac{\hat\mu^{th} - \hat\mu^{exp}}{\delta\mu}\right)^2$. For other channels where we know the correlations between the rates for $ggH$/$ttH$ or VBF/VH production modes, we use the experimental 2D likelihood functions $\chi_{2D}^2$. For electroweak precision data the correlations are known and enter into the fit via the chi-squared function $\chi_{EWPT}^2$~\cite{Falkowski:2013dza,Falkowski:private_comm_ChiEWPT}. The fitting procedure then consists in minimizing the following $\chi^2$-function:
	\begin{equation*}
		\chi^2(\{c_i, \widetilde{c}_i\}) = \chi_{EWPT}^2(\{c_i\}) + \sum \chi_{1D}^2(\hat\mu^{th}, \hat\mu^{exp} \pm \delta\mu) + \sum \chi_{2D}^2(\hat\mu_{ggH+ttH}^{th}, \hat\mu_{VBF+VH}^{th}) + \chi_\lambda^2(\lambda \pm \delta\lambda) \, .
	\end{equation*}
	In the Gaussian approximation the $\chi_{EWPT}^2$ function can be approximated around its best-fit point $(c_V^0, c_{\gamma\gamma}^0, c_{Z\gamma}^0)$ by the following quadratic form:
	\begin{equation}
		\chi_{EWPT}^2(\{c_i\}) = 193.005 + \sum_{i,j = V, \gamma\gamma, Z\gamma} (c_i - c_i^0) (\sigma^2)^{-1}_{ij} (c_j - c_j^0) \, , \quad \text{where:} \quad (\sigma^2)_{ij} = \sigma_i \rho_{ij} \sigma_j ,
	\end{equation}
	its minimum point with the corresponding $1\sigma$ deviations $\sigma_i$ for each component $\{c_V, c_{\gamma\gamma}, c_{Z\gamma}\}$, and the correlation matrix, being:
	\begin{equation}
		\begin{aligned}
			c_V^0 &= 1.082 \pm 0.066 \\
			c_{\gamma\gamma}^0 &= 0.096 \pm 0.653 \\
			c_{Z\gamma}^0 &= -0.036 \pm 0.915
		\end{aligned}
		\quad ; \quad
		\rho =
			\begin{pmatrix}
				 1     &  0.275 & -0.138 \\
				 0.275 &  1     & -0.989 \\
				-0.138 & -0.989 & 1
			\end{pmatrix}
		.
	\end{equation}

	\par We also incorporate in the fit the large uncertainty on the prediction of the SM $ggH$ production cross-section by introducing a nuisance parameter $\lambda$ with a Gaussian distribution around the central value, via the $\chi_\lambda^2$ term: for the LHC at $\sqrt{s}=8$~TeV we take~\cite{Heinemeyer:2013tqa} the scale error: ($+7.2\%$, $-7.8\%$) and the PDF error: ($+7.5\%$, $-6.9\%$) and add those two linearly. The treatment of such theoretical uncertainties in Higgs fits is extensively reviewed in~\cite{Fichet:2015xla}.

	\subsection{Fit over the CP-even Parameters}
		\par The 7 CP-even parameters (Eq.~\ref{eq:CPEvenParameters}) are fitted to the available Higgs and electroweak precision data, while fixing the CP-odd ones to zero. The central values and $68\%$~CL intervals for the parameters are summarized in Table~\ref{tab:fit_cpeven}.


		\begin{table}[!htbp]
			\centering
			\caption{Fit results for CP-even parameters, given as central values with $1\sigma$~CLs, obtained in Gaussian approximation (left) and after marginalization over the remaining parameters (right).}
			\label{tab:fit_cpeven}

			\renewcommand*{\arraystretch}{1.3} 

			\begin{tabular}{ c | c | c | }
				\cline{2-3}
					& Gaussian & Marginalized \\
				\hline
				\multicolumn{1}{|c|}{$c_V$}
					& $1.034 \pm 0.023$ & $1.034^{+0.023}_{-0.030}$ \\
				\multicolumn{1}{|c|}{$c_u$}
					& $1.467 \pm 0.203$ & $1.467^{+0.179}_{-0.120}$ \\
				\multicolumn{1}{|c|}{$c_d$}
					& $0.811 \pm 0.136$ & $0.961^{+0.144}_{-0.158}$ \\
				\multicolumn{1}{|c|}{$c_l$}
					& $0.941 \pm 0.129$ & $0.944^{+0.132}_{-0.133}$ \\
				\multicolumn{1}{|c|}{$c_{gg}$}
					& $-0.0063 \pm 0.0025$ & $-0.0062^{+0.0023}_{-0.0026}$ \\
				\multicolumn{1}{|c|}{$c_{\gamma\gamma}$}
					& $0.0007 \pm 0.0009$ & $-0.0003 \pm 0.0004$ \\
				\multicolumn{1}{|c|}{$c_{Z\gamma}$}
					& $0.004 \pm 0.015$ & $0.004^{+0.013}_{-0.040}$ \\
				\hline
			\end{tabular}
		\end{table}

		\par We find a $\Delta\chi^2 = \chi_{\rm SM}^2 - \chi_{\rm min}^2 = 10.2$, meaning that the SM gives a correct fit to the Higgs and electroweak precision data. When quoting the confidence regions above we ignored the degenerate minima of the likelihood function isolated from the SM point where a large 2-derivative Higgs coupling conspires with the SM loop contributions to produce a small shift of the Higgs observables. The current data already put meaningful limits on {\em all} 7 parameters. The strong constraint on $c_V$ is dominated by electroweak precision data, and ignoring them in the fit weakens the constraint, and one obtains $c_V = 0.967^{+0.088}_{-0.105}$. It can also be relaxed in the presence of additional tuned contributions to the $S$ and $T$ parameters that could arise from integrating out heavy new physics states.

		\par
\begin{figure}[h!]
			\centering
			\subfloat[]{\label{subfig:cu_cgg}   \includegraphics[width=0.4\textwidth]{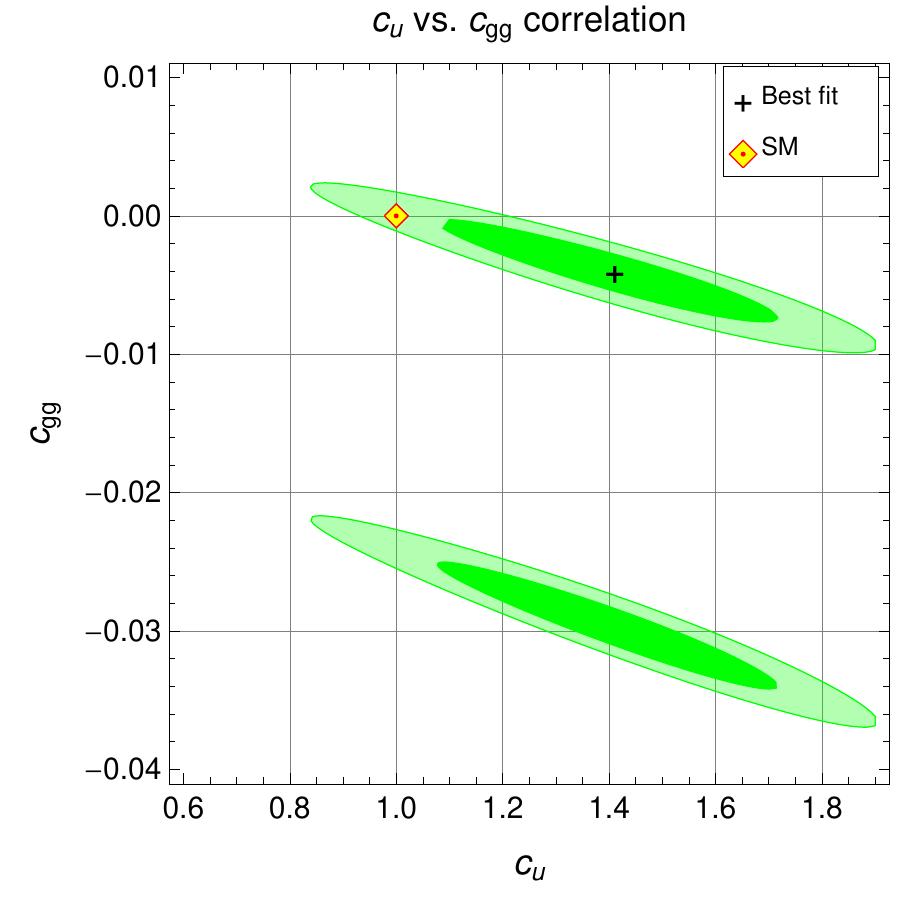}} \hspace{2cm}
			\subfloat[]{\label{subfig:cut_cggt} \includegraphics[width=0.4\textwidth]{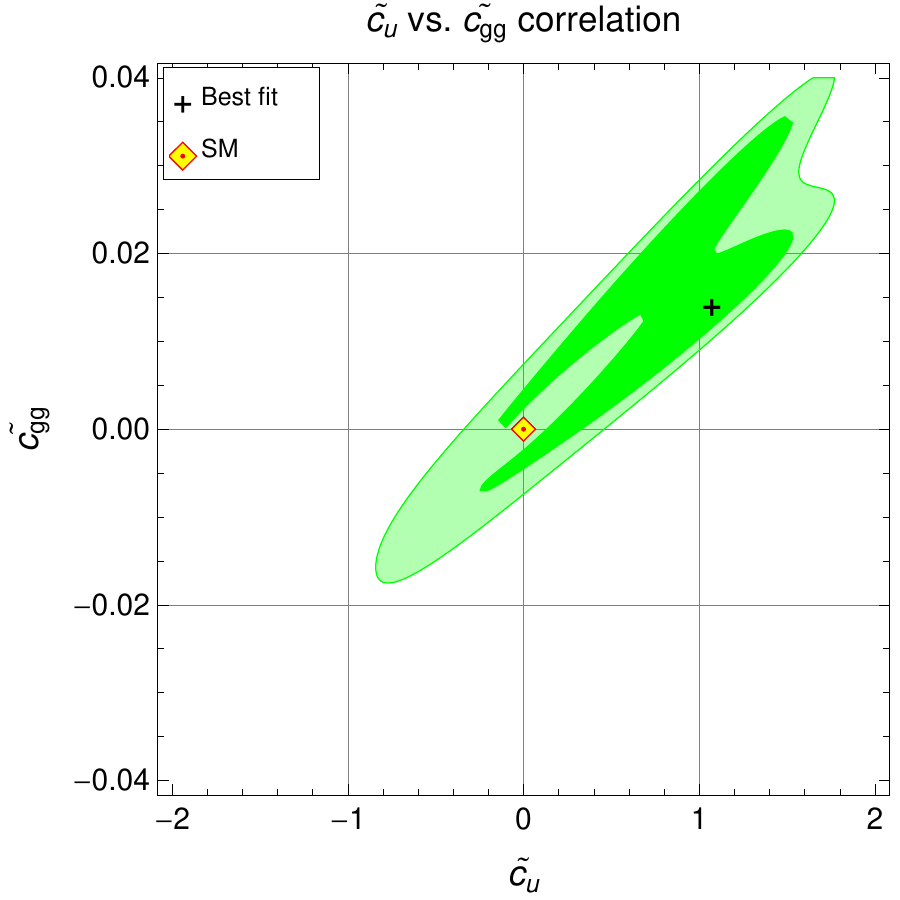}}
			\caption{(Left) Fit in the $c_u$--$c_{gg}$ plane with the other couplings set to their best-fit central values. (Right) Fit in the $\widetilde{c}_u$--$\widetilde{c}_{gg}$ plane with the other couplings set to their best-fit central values. Dark green: $68\%$~CL; light green: $95\%$~CL. See details in the text.}
			\label{fig:cu_cgg}
		\end{figure}

 The fit features an approximately flat correlation region for the $c_{gg}$ and $c_u$ couplings, corresponding to the combination that sets the strength of the gluon fusion production mode. This is clearly visible in Fig.~\ref{subfig:cu_cgg} where a 2D fit in the $c_u$--$c_{gg}$ plane is performed, whereas the other couplings are set to their best-fit central values. The results~\cite{Aad:2015gra,Khachatryan:2015ila, Aad:2015iha,Khachatryan:2014qaa} of ATLAS and CMS in the $ttH$ production channel, which depend on $c_u$ only, provide interesting constraints on $c_u$ independently of $c_{gg}$. The fit shows also a strong preference for $c_d \neq 0$ even though the $h \to b \bar{b}$ decay has not been clearly observed. The reason is that $c_d$ determines $\Gamma_{bb}$ which dominates the total Higgs decay width and the latter is indirectly constrained by the Higgs rates measured in other decay channels.

		\par Concerning loop-generated effective couplings, the least stringent constraint is currently the one on $c_{Z\gamma}$ which reflects the weak experimental limits on the $h \to Z\gamma$ decay rate. There are good prospects~\cite{Chen:2013ejz,Chen:2014pia,Chen:2014gka} of probing $c_{Z\gamma}$, as well as $c_{\gamma\gamma}$, using differential cross-section measurements in the "Golden Channel" $h \to 4\ell$.

	\subsection{Global fit over the CP-even and CP-odd Parameters}
		\par In this section both CP-even and CP-odd parameters (Eq.~\ref{eq:CPOddParameters}) are fitted together. We make the same assumptions about data errors as in the case of the CP-even fit and we continue to take into account the large uncertainty in the prediction of the SM $ggH$ production cross-section. The central values and $68\%$~CL intervals for the parameters are summarized in Table~\ref{tab:fit_cpevenodd}.

		\begin{table}[!htbp]
			\centering
			\caption{Fit results for CP-even and CP-odd parameters, given as central values with $1\sigma$~CLs, obtained in Gaussian approximation (left) and after marginalization over the remaining parameters (right).}
			\label{tab:fit_cpevenodd}

			\renewcommand*{\arraystretch}{1.3} 

			\begin{tabular}{ c c | c | c | }
				\cline{3-4}
					&& Gaussian & Marginalized \\
				\hline
				\multicolumn{1}{|c}{\multirow{7}{*}{CP-even}}
					& $c_V$ & $1.028 \pm 0.024$ & $1.031^{+0.023}_{-0.024}$ \\
				\multicolumn{1}{|c}{}
					& $c_u$ & $0.983 \pm 0.395$ & $1.464^{+0.197}_{-0.172}$ \\
				\multicolumn{1}{|c}{}
					& $c_d$ & $0.977 \pm 0.200$ & $0.836^{+0.133}_{-0.122}$ \\
				\multicolumn{1}{|c}{}
					& $c_l$ & $1.003 \pm 0.149$ & $1.010^{+0.112}_{-0.116}$ \\
				\multicolumn{1}{|c}{}
					& $c_{gg}$ & $-0.0184 \pm 0.0075$ & $-0.0015^{+0.0026}_{-0.0063}$ \\
				\multicolumn{1}{|c}{}
					& $c_{\gamma\gamma}$ & $-0.0013 \pm 0.0029$ & $-0.0013^{+0.0021}_{-0.0040}$ \\
				\multicolumn{1}{|c}{}
					& $c_{Z\gamma}$ & $0.0025 \pm 0.0193$ & $0.0025^{+0.0118}_{-0.0314}$ \\
				\hline
				\multicolumn{1}{|c}{\multirow{6}{*}{CP-odd}}
					& $\widetilde{c}_u$ & $0.008 \pm 0.354$ & $\cdots$ \\
				\multicolumn{1}{|c}{}
					& $\widetilde{c}_d$ & $0.027 \pm 0.432$ & $-0.003^{+0.328}_{-0.335}$ \\
				\multicolumn{1}{|c}{}
					& $\widetilde{c}_l$ & $0.035 \pm 1.156$ & $0.035^{+0.483}_{-0.573}$ \\
				\multicolumn{1}{|c}{}
					& $\widetilde{c}_{gg}$ & $0.0105 \pm 0.0045$ & $\cdots$ \\
				\multicolumn{1}{|c}{}
					& $\widetilde{c}_{\gamma\gamma}$ & $-0.0047 \pm 0.0041$ & $0.0041^{+0.0055}_{-0.0138}$ \\
				\multicolumn{1}{|c}{}
					& $\widetilde{c}_{Z\gamma}$ & $-0.0027 \pm 0.0670$ & $-0.0027^{+0.0303}_{-0.0237}$ \\
				\hline
			\end{tabular}
		\end{table}

		\par
\begin{figure}[h!]
			\centering
			\subfloat{\includegraphics[width=0.4\textwidth]{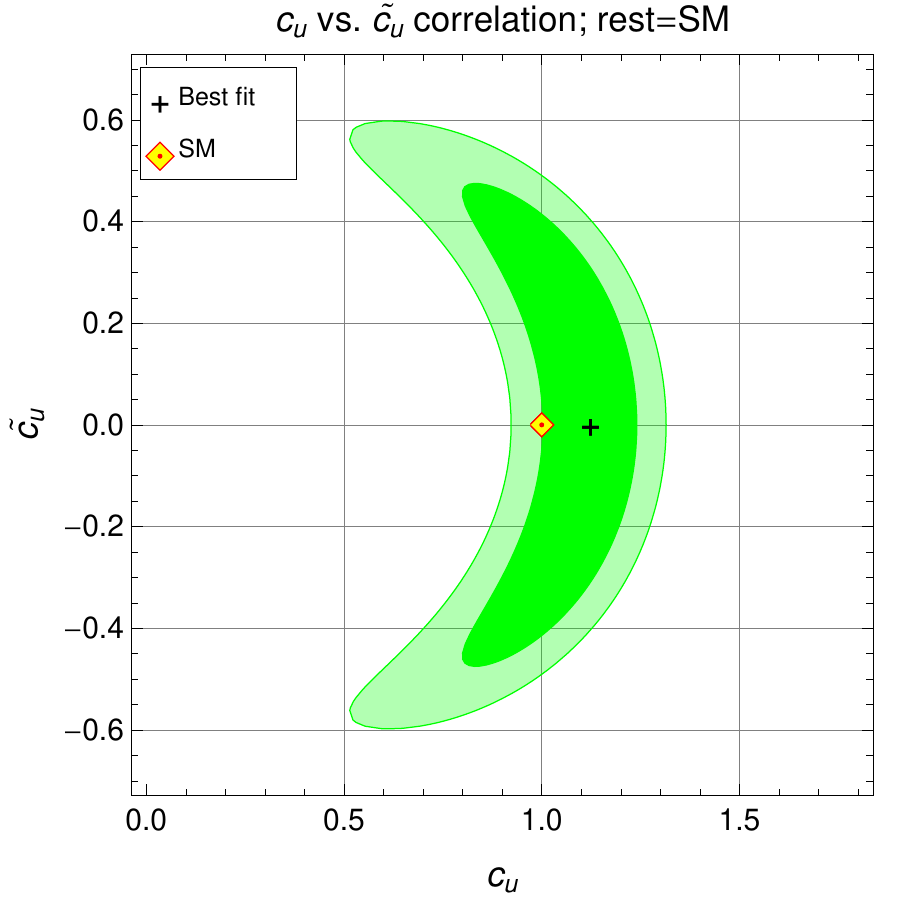}} \hspace{2cm}
			\subfloat{\includegraphics[width=0.4\textwidth]{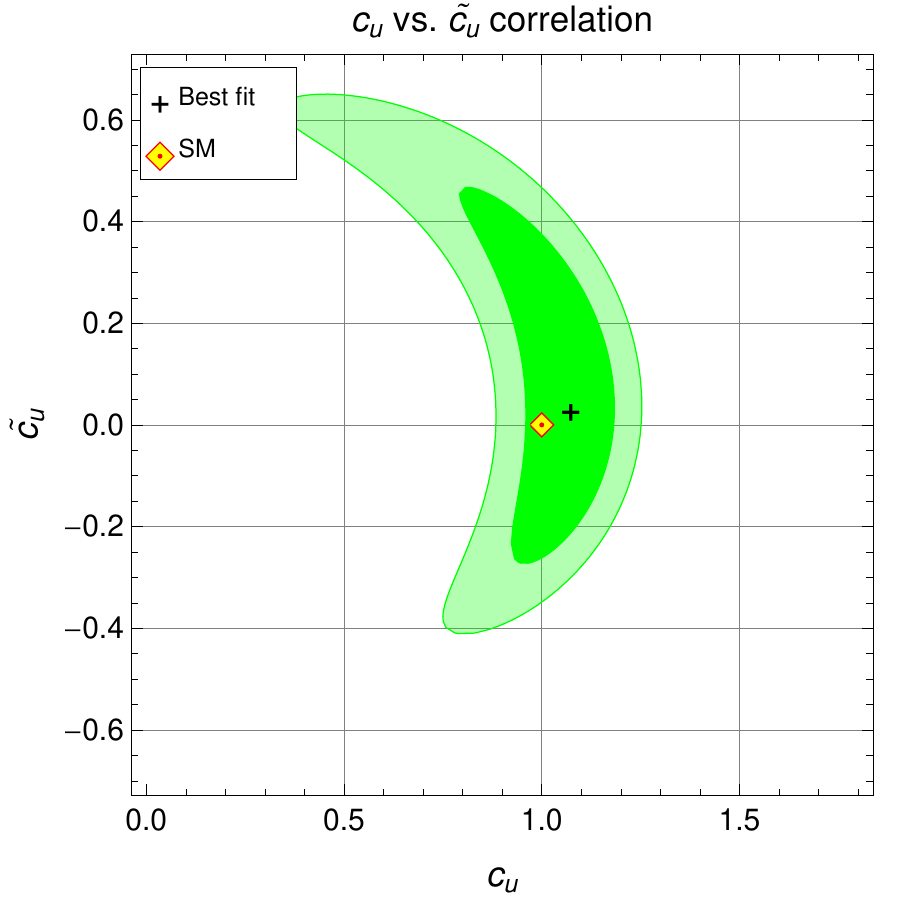}}
			\caption{A fit in the $c_u$--$\widetilde{c}_u$ plane, with the other couplings fixed to their Standard Model values (Left), or set to their best-fit central values (Right) in the Gaussian approximation, but with $c_{gg} = 0 = \widetilde{c}_{gg}$. Dark green: $68\%$~CL; light green: $95\%$~CL. See details in the text.}
			\label{fig:cu_cut}
		\end{figure}

\begin{figure}[h!]
			\centering
			\subfloat{\includegraphics[width=0.4\textwidth]{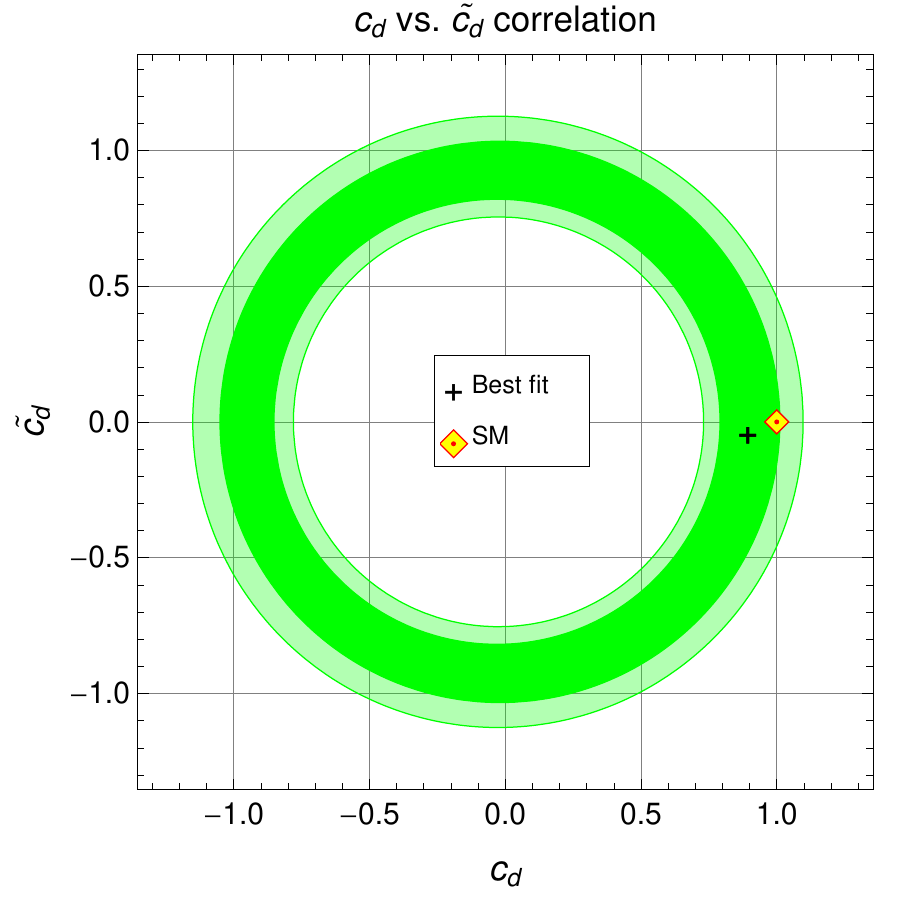}} \hspace{2cm}
			\subfloat{\includegraphics[width=0.4\textwidth]{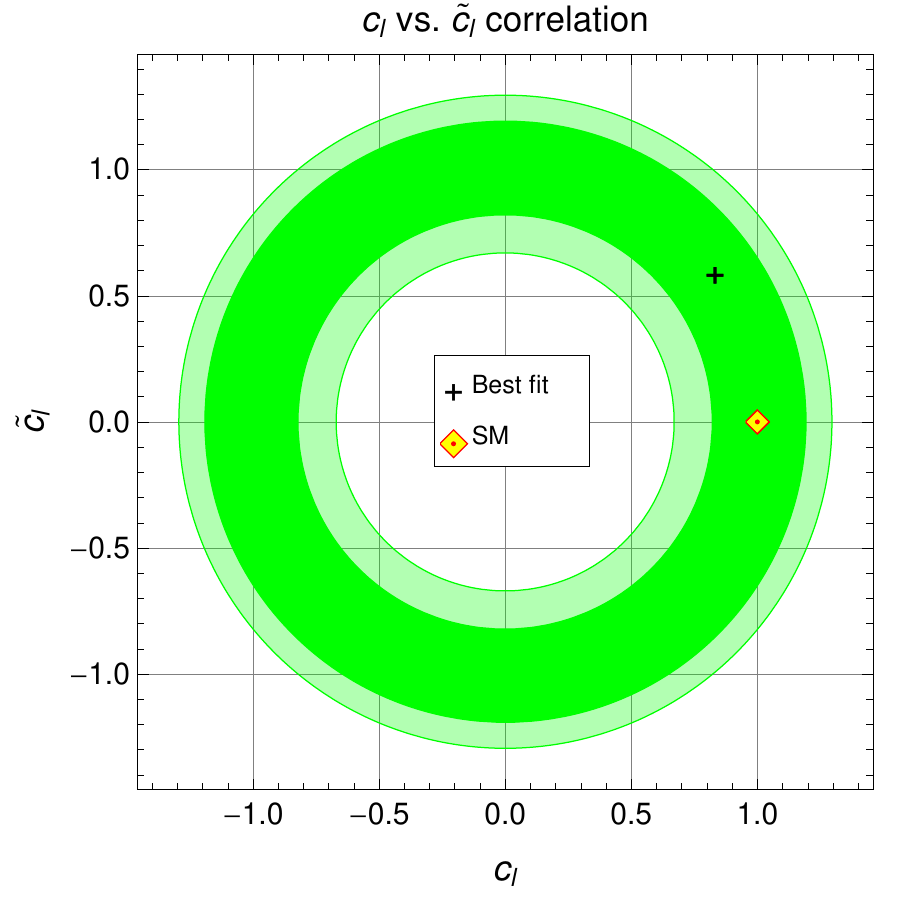}}
			\caption{Fits in the $c_d$--$\widetilde{c}_d$ plane (Left), and in the $c_l$--$\widetilde{c}_l$ plane (Right), with the other couplings set to their best-fit central values. Current Higgs signal rates do not significantly constrain the signs of the couplings. The displayed best-fit points are compatible with the large $1\sigma$ errors for the CP-odd couplings found in the global fit.}
			\label{fig:cdcl_cdtclt}
		\end{figure}

 A value of $\Delta\chi^2 = \chi_{\rm SM}^2 - \chi_{\rm min}^2 = 4.02$ is obtained. We note that while the CP-odd Higgs-gauge couplings are globally constrained by current data, the CP-odd fermionic (up/down-type and leptonic) couplings have large $1\sigma$ errors, meaning their sign is not constrained (see also Figs.~\ref{fig:cu_cut} and~\ref{fig:cdcl_cdtclt}). This is due to the fact that the Higgs rate measurements from the LHC mostly constrain the sum of the squares of the CP-even and CP-odd couplings (see Section~\ref{subsec:RelDecWidth}) in the fermionic sector.

		\par As in the CP-even case, this fit also features a flat correlation region for the $\widetilde{c}_{gg}$ and $\widetilde{c}_u$ couplings, see Fig.~\ref{subfig:cut_cggt}. The CP-even and odd couplings $c_{gg}$, $\widetilde{c}_{gg}$ and $c_u$, $\widetilde{c}_u$ are in competition because their combination sets the strength of the gluon fusion production mode, and $c_{gg}$ and $\widetilde{c}_{gg}$ are only constrained by $ggH$ production. We also see that if those couplings are allowed to float, the fit would prefer an $\mathcal{O}(1)$ positive value for $\widetilde{c}_u$ and a non-zero $\widetilde{c}_{gg}$, while driving $c_u$ towards larger values, whereas the SM point is still compatible at $68\%$~CL level. In Fig.~\ref{fig:cu_cut} we show the correlation regions of the $c_u$ and $\widetilde{c}_u$ couplings when the other couplings are set to their SM values, and when they are set to their best-fit central values in the Gaussian approximation, but with both $c_{gg}$ and $\widetilde{c}_{gg}$ set to zero. Setting $c_{gg}$ and $\widetilde{c}_{gg}$ to their central values would move the $68\%$ and $95\%$~CL regions far away from the SM point. The fit appears to be a bit sensitive to the sign of $\widetilde{c}_u$ when using the best-fit values for the other couplings, due to loop contributions from $h \to \gamma\gamma$ and $h \to Z\gamma$.

		\par To break the sign degeneracy of the fermionic CP-odd couplings and improve the precision on their determination, other types of studies are needed. A first one is to study differential cross-section measurements, for example in the "Golden Channel"~\cite{Chen:2013ejz,Chen:2014pia,Chen:2014gka}, or via jet kinematics in the VBF~\cite{Djouadi:2013yb} or in VH~\cite{Godbole:2013saa} production modes. Alternatively for the up-type coupling, methods involving mass distributions as well as top-quark polarization and spin correlations can be done in the $t\bar{t}H$, $tH$ and $\bar{t}H$ production channels~\cite{Ellis:2013yxa}. A second one is to use EDMs as shown by J.Brod et al.~\cite{Brod:2013cka}: assuming that the Higgs couples to the first generation of fermions with SM couplings, constraints on the $c_i$ and $\widetilde{c}_i$ couplings can be derived for the top and bottom quarks and tau lepton by using low-energy bounds on the EDMs of the electron and the neutron together with existing Higgs production data. It is shown that those limits can be dramatically enhanced if bounds on EDMs are improved from a factor of 100 to 300, from $|d_e/e| < \SI{8.7e-29}{cm}$ to $< \SI{e-30}{cm}$ for the electron and $|d_n/e| < \SI{2.9e-26}{cm}$ to $< \SI{e-28}{cm}$ for the neutron, while using \SI{3000}{fb^{-1}} of updated Higgs data from the \SI{14}{TeV} high-luminosity LHC upgrade. With these expected improvements the following limits are obtained: $c_t = 1.00 \pm 0.03$ and $\widetilde{c}_t = 0.00 \pm \SI{2e-4}{}$ (other couplings fixed to their SM values), and $c_b = 1.00 \pm 0.08$ and $\widetilde{c}_b = 0.00 \pm 0.02$. If the assumption that the Higgs couples to the first generation with SM couplings is removed, then constraints from the neutron EDM can be still used, and improving its bounds may still allow to constrain $\widetilde{c}_t$ for the top quark at the same level as before. We refer to their paper~\cite{Brod:2013cka} for an extended discussion. Concerning Higgs leptonic couplings, electron EDM reduces the possibility for large values of the CP-odd $\tau$ lepton coupling $\widetilde{c}_\tau$, of order $0.01$, while keeping a sign degeneracy on its CP-even coupling $c_\tau$.

		\par The constraints on the loop-generated CP-odd effective couplings $\widetilde{c}_{\gamma\gamma}$ and $\widetilde{c}_{Z\gamma}$ may be improved~\cite{Chen:2014ona} by looking for a possible forward-backward asymmetry of charged leptons in the 3-body decay $h \to \ell^- \ell^+ \gamma$.

\sectionnn{Summary}
	\par In this work we employed an effective theory approach for parametrizing small deviations of Higgs couplings to matter from their SM prediction, by using a phenomenological Lagrangian. To derive it we started from the dimension-6 SILH Lagrangian of Giudice {\it et~al.}~\cite{Giudice:2007fh} written in the basis employed by Contino {\it et~al.}~\cite{Contino:2013kra}, where extra custodial relations relate some of the effective Wilson coefficients together. Since they introduce power divergences in the oblique parameters at loop-level, we argue that, due to current EW constraints, those divergences must disappear. Making the hypothesis that there is no accidental cancellations between operators of different types, namely the CP-even, CP-odd and the $\kappa_i$ ones, allows us to obtain extra constraints on the parameters of the theory that reduce the number of free parameters of the phenomenological Lagrangian: 7 parameters on the CP-even sector and 6 parameters on the CP-odd sector of the theory are obtained. They are then fitted to current Higgs data and electroweak precision measurements.

	\par Using the current LHC Higgs rates we are able to constrain CP-even parameters and some CP-odd ones. However one should note that, since until now the rate measurements almost only constrain the sum of the squares of the CP-even and odd couplings of the Higgs boson in the fermionic sector, they are constrained only via their absolute values and are degenerated in sign; therefore more elaborate methods are needed to constrain their possible values and break the sign degeneracies.

	\par So far no indication for large deviations from the SM are found in the fits since all of the fitted parameters are compatible with their SM values withing $68\%$~CL.

	\sectionnn{Acknowledgements}
		\par I thank Adam Falkowski and Ulrich Ellwanger for guidance through this work and helpful remarks. I also thank the CMS group of CEA Saclay, for discussions about the CMS results and giving me the opportunity to present my work.

	\noindent\hrulefill

%
%
%

\appendix

\section{Standard Model Lagrangian -- Conventions}
\label{subsec:SMLagr}

	\par In this paper we use the convention of~\cite{Contino:2013kra}, so that the Standard Model Lagrangian writes:
	\begin{equation}
		\mathcal{L}_\text{SM} = - \frac{1}{4} \sum_V V^a_{\mu\nu} V^{a{\mu\nu}} + \overline{f_L}^i \imath \centernot{D} f_L^i + \overline{f_R}^i \imath \centernot{D} f_R^i + \left(Y_{ij} \overline{f^i_L} H f^j_R + \hc\right) + \left| D_\mu H \right|^2 - V(H) \; .
	\end{equation}
	We understand implicit summation over repeated indices, where $(i,j) = 1,2,3$ span the three lepton families. For each gauge group $U(1)_Y$, $SU(2)_L$ and $SU(3)_C$ correspond the vector fields $V = B, W^{a = 1,2,3}, G^{a = 1 \dots 8}$ respectively and we denote by $t_V^a$ and $g_V$ the corresponding generators and gauge couplings for each group. Their field-strength tensors write:
	\begin{equation}
		V^a_{\mu\nu} = \partial_\mu V^a_\nu - \partial_\nu V^a_\mu + g_V f^{abc} V^b_\mu V^c_\nu
	\end{equation}
	in general, with $f^{abc}$ the corresponding structure constants of the gauge groups. For $U(1)$ the self-iteracting term is absent because the group is abelian. The covariant derivative is:
	\begin{equation}
		D_\mu = \partial_\mu - \imath \sum_V g_V t_V^a V^a_\mu
	\end{equation}
	that acts on the different fields according to their quantum numbers (see~\cite{Contino:2013kra}). The two last terms of the Lagrangian are the kinetic term and the potential of the Higgs field,
	\begin{equation}
		V(H) = -\mu_H^2 H^\dagger H + \lambda (H^\dagger H)^2
	\end{equation}
	where $\mu_H^2 = 2 \lambda v^2$, $v$ being the Higgs vev when it develops a non-trivial minima region, leading to EWSB.

\section{One-loop corrections for the $S$, $T$, $U$ parameters}

\subsection{Corrections to the gauge bosons propagators}
\label{subsec:1loop}

	\begin{figure}[h!]
		\centering
		\includegraphics[width=0.3\textwidth]{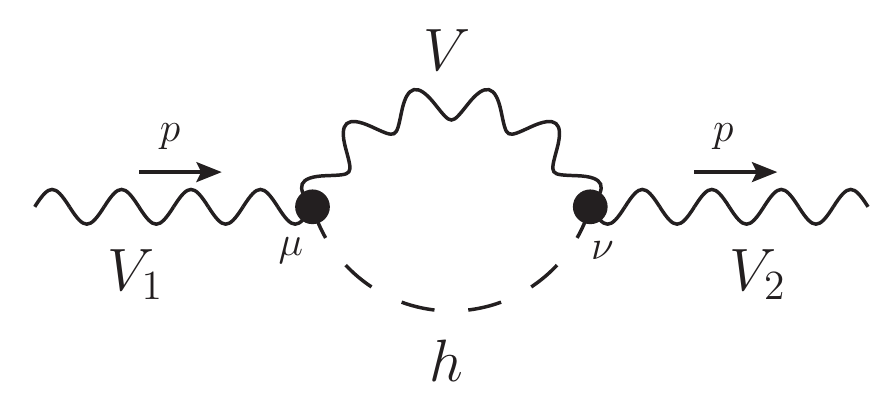}
		\caption{The one-loop correction $V_1$--($V$/$H$)--$V_2$ with one Higgs field.}
	\end{figure}

	\par Considering only linear Higgs couplings in this work, the evaluated one-loop corrections due to the effective Lagrangian present only one Higgs propagator.

	\par We use a hard cut-off scheme ($\Lambda \rightarrow +\infty$) for computing the loop integral and we keep its divergence up to the $\ln{\Lambda^2}$ order. We define $\widetilde{\Lambda} \equiv \frac{\Lambda}{M}$, $M$ being the EW scale to write a dimensionless quantity inside the logarithm. Now, after defining an auxiliary function $f$ and the projector $P^2$ as:
	\begin{align}
		f(\Lambda^2, p^2, m_V^2) &= \frac{1}{16 \pi^2 v^2} \left[\Lambda^2 - \ln{\widetilde{\Lambda}^2} \left(m_H^2 + m_V^2 - \frac{p^2}{3}\right)\right] \, , \\
		P^2_{\mu\nu} &= \left(p^2 g_{\mu\nu} - p_\mu p_\nu\right) \, ,
	\end{align}
	the analytical expressions of the corrections up to order $\ln{\widetilde{\Lambda}^2}$ write:

	\begin{itemize}[leftmargin=*]
		\item[\textbullet] \uline{$Z$--$Z$ correction}:
			\par $Z$--($Z$/$H$)--$Z$ loop:
			\begin{adjustwidth}{-1cm}{}\resizemath{\small}

			\begin{equation}\begin{split}
				\Pi_{\mu\nu} = & \left[ \frac{-3 \kappa_{ZZ}^2}{8} \frac{\Lambda^4}{16 \pi^2 v^2} - \left[ c_Z^2 m_Z^2 + 3 c_Z \kappa_{ZZ} m_Z^2 - \frac{3 \kappa_{ZZ}^2}{4} \left(m_H^2 + m_Z^2\right) \right] f(\Lambda^2, p^2, m_Z^2) + \frac{3 \kappa_{ZZ}^2}{4} m_H^2 \left(m_Z^2 - \frac{p^2}{3}\right) \frac{\ln{\widetilde{\Lambda}^2}}{16 \pi^2 v^2} \right] g_{\mu\nu} \\
				&- 4 c_Z^2 m_Z^2 \left(m_Z^2 g_{\mu\nu} - \frac{p_\mu p_\nu}{3}\right) \frac{\ln{\widetilde{\Lambda}^2}}{16 \pi^2 v^2} + \left[ \left[ \frac{c_{ZZ}^2 - \widetilde{c}_{ZZ}^2}{2} + c_Z \kappa_{ZZ} + \frac{3 c_{ZZ} \kappa_{ZZ}}{2} + \frac{\kappa_{ZZ}^2}{12} \left(26 - \frac{3 p^2}{m_Z^2}\right) \right] f(\Lambda^2, p^2, m_Z^2) \right. \\
				&\left. + \frac{1}{3} \left[ c_Z \left(6 c_{ZZ} + 16 \kappa_{ZZ}\right) m_Z^2 - \left(c_{ZZ}^2 + \frac{8 \kappa_{ZZ}^2}{3}\right) p^2 - \left(\frac{3 c_{ZZ}}{2} + \kappa_{ZZ}\right) \kappa_{ZZ} \left(m_H^2 - m_Z^2 + 3 p^2\right) \right] \frac{\ln{\widetilde{\Lambda}^2}}{16 \pi^2 v^2} \right] P^2_{\mu\nu}
			\end{split}\end{equation}

			\endresizemath\end{adjustwidth}

			$Z$--($\gamma$/$H$)--$Z$ loop:
			\begin{adjustwidth}{-1cm}{}\resizemath{\small}
			\begin{equation}\begin{split}
				\Pi_{\mu\nu} = & \frac{\kappa_{Z\gamma}^2}{4} \left[ \frac{-3}{2} \frac{\Lambda^4}{16 \pi^2 v^2} + 3 m_H^2 f(\Lambda^2, p^2, 0) - m_H^2 p^2 \frac{\ln{\widetilde{\Lambda}^2}}{16 \pi^2 v^2} \right] g_{\mu\nu} + \left[ \left[ \frac{c_{Z\gamma}^2 - \widetilde{c}_{Z\gamma}^2}{2} + \frac{3 c_{Z\gamma} \kappa_{Z\gamma}}{2} + \frac{2 \kappa_{Z\gamma}^2}{3} \right] f(\Lambda^2, p^2, 0) \right.\\
				&\left. - \left[ \left(\frac{c_{Z\gamma}}{2} + \frac{\kappa_{Z\gamma}}{3}\right) \kappa_{Z\gamma} \left(m_H^2 + \frac{2 p^2}{3}\right) + c_{Z\gamma} \left(c_{Z\gamma} + \frac{\kappa_{Z\gamma}}{2}\right) \frac{p^2}{3} \right] \frac{\ln{\widetilde{\Lambda}^2}}{16 \pi^2 v^2} \right] P^2_{\mu\nu}
			\end{split}\end{equation}
			\endresizemath\end{adjustwidth}

		\item[\textbullet] \uline{$W$--$W$ correction}: There is only the $W$--($W$/$H$)--$W$ loop:
		\begin{adjustwidth}{-2cm}{}\resizemath{\small}

			\begin{equation}\begin{split}
				\Pi_{\mu\nu} = & \left[ \frac{-3 \kappa_{WW}^2}{8} \frac{\Lambda^4}{16 \pi^2 v^2} - \left[c_W^2 m_W^2 + 3 c_W \kappa_{WW} m_W^2 - \frac{3 \kappa_{WW}^2}{4} \left(m_H^2 + m_W^2\right)\right] f(\Lambda^2, p^2, m_W^2) + \frac{3 \kappa_{WW}^2}{4} m_H^2 \left(m_W^2 - \frac{p^2}{3}\right) \frac{\ln{\widetilde{\Lambda}^2}}{16 \pi^2 v^2} \right] g_{\mu\nu} \\
				&- 4 c_W^2 m_W^2 \left(m_W^2 g_{\mu\nu} - \frac{p_\mu p_\nu}{3}\right) \frac{\ln{\widetilde{\Lambda}^2}}{16 \pi^2 v^2} + \left[ \frac{c_{WW}^2 - \widetilde{c}_{WW}^2}{2} + c_W \kappa_{WW} + \frac{3 c_{WW} \kappa_{WW}}{2} + \frac{\kappa_{WW}^2}{12} \left(26 - \frac{3 p^2}{m_W^2}\right) \right] f(\Lambda^2, p^2, m_W^2) P^2_{\mu\nu} \\
				&+ \frac{1}{3} \left[ c_W \left(6 c_{WW} + 16 \kappa_{WW}\right) m_W^2 - \left(c_{WW}^2 + \frac{8 \kappa_{WW}^2}{3}\right) p^2 - \left(\frac{3 c_{WW}}{2} + \kappa_{WW}\right) \kappa_{WW} \left(m_H^2 - m_W^2 + 3 p^2\right) \right] \frac{\ln{\widetilde{\Lambda}^2}}{16 \pi^2 v^2} P^2_{\mu\nu}
			\end{split}\end{equation}

			\endresizemath\end{adjustwidth}

		\item[\textbullet] \uline{$\gamma$--$\gamma$ correction}:
			\par $\gamma$--($\gamma$/$H$)--$\gamma$ loop:
			\resizemath{\small}
			\begin{equation}
				\Pi_{\mu\nu} = \left[ \frac{c_{\gamma\gamma}^2 - \widetilde{c}_{\gamma\gamma}^2}{2} f(\Lambda^2, p^2, 0) - c_{\gamma\gamma}^2 \frac{p^2}{3} \frac{\ln{\widetilde{\Lambda}^2}}{16 \pi^2 v^2} \right] P^2_{\mu\nu}
			\end{equation}
			\endresizemath

			$\gamma$--($Z$/$H$)--$\gamma$ loop:
			\resizemath{\small}
			\begin{equation}
				\Pi_{\mu\nu} = \left[ \left( \frac{c_{Z\gamma}^2 - \widetilde{c}_{Z\gamma}^2}{2} - \kappa_{Z\gamma}^2 \frac{p^2}{4 m_Z^2} \right) f(\Lambda^2, p^2, m_Z^2) - \left(c_{Z\gamma}^2 + 3 c_{Z\gamma} \kappa_{Z\gamma} + 3 \kappa_{Z\gamma}^2\right) \frac{p^2}{3} \frac{\ln{\widetilde{\Lambda}^2}}{16 \pi^2 v^2} \right] P^2_{\mu\nu}
			\end{equation}
			\endresizemath

		\item[\textbullet] \uline{$Z \rightarrow \gamma$ mixing}:
			\par $Z$--($Z$/$H$)--$\gamma$ loop:
			\vspace{-0.5cm}
			\begin{adjustwidth}{-1cm}{}\resizemath{\small}
			\begin{equation}\begin{split}
				\Pi_{\mu\nu} = & \left[ \left[ \frac{c_{ZZ} c_{Z\gamma} - \widetilde{c}_{ZZ} \widetilde{c}_{Z\gamma}}{2} + \frac{c_Z \kappa_{Z\gamma}}{2} + \frac{3}{4} \kappa_{ZZ} \left( c_{Z\gamma} + \kappa_{Z\gamma} \right) - \kappa_{ZZ} \kappa_{Z\gamma} \frac{p^2}{4 m_Z^2} \right] f(\Lambda^2, p^2, m_Z^2) + \left[\vphantom{\frac{p^2}{6}} c_Z (c_{Z\gamma} + 2 \kappa_{Z\gamma}) m_Z^2 \right.\right. \\
				&\left.\left. - \frac{c_{Z\gamma} \kappa_{ZZ}}{4} (m_H^2 - m_Z^2 + 3 p^2) - \left(2 c_{ZZ} c_{Z\gamma} + 3 c_{ZZ} \kappa_{Z\gamma} + 8 \kappa_{ZZ} \kappa_{Z\gamma}\right) \frac{p^2}{6} \right] \frac{\ln{\widetilde{\Lambda}^2}}{16 \pi^2 v^2} \right] P^2_{\mu\nu}
			\end{split}\end{equation}
			\endresizemath\end{adjustwidth}

			$Z$--($\gamma$/$H$)--$\gamma$ loop:
			\begin{adjustwidth}{-1cm}{}\resizemath{\small}
			\begin{equation}
				\Pi_{\mu\nu} = \left[ \left[ \frac{c_{Z\gamma} c_{\gamma\gamma} - \widetilde{c}_{Z\gamma} \widetilde{c}_{\gamma\gamma}}{2} + \frac{3 c_{\gamma\gamma} \kappa_{Z\gamma}}{4} \right] f(\Lambda^2, p^2, 0) - c_{\gamma\gamma} \left[ c_{Z\gamma} \frac{p^2}{3} + \frac{\kappa_{Z\gamma}}{4} \left(m_H^2 + p^2\right) \right] \frac{\ln{\widetilde{\Lambda}^2}}{16 \pi^2 v^2} \right] P^2_{\mu\nu}
			\end{equation}
			\endresizemath\end{adjustwidth}
	\end{itemize}

\subsection{Corrections to $S$, $T$, $U$}
\label{subsec:STUCorr}

	\par Using our previous results in the Peskin-Takeuchi parameters, using: $f(\Lambda^2, 0, 0) = \frac{1}{16 \pi^2 v^2} \left(\Lambda^2 - m_H^2 \ln{\widetilde{\Lambda}^2}\right)$ and defining for a given vector boson $V$: ${\epsilon(V) = \left(c_V^2 - 1\right) - 6 c_V c_{VV} - 10 c_V \kappa_{VV} + 3 c_{VV} \kappa_{VV} + \frac{19 \kappa_{VV}^2}{4}}$, we are led to the following expressions:

	\begin{adjustwidth}{-1cm}{}\resizemath{\small}
	\vspace*{-0.5cm}
	\begin{align}
		\begin{split}
		\alpha S =
			&\; 2 s_w^2 c_w^2 f(\Lambda^2, 0, 0) \left[
			\begin{aligned}
				&c_{ZZ}^2 - c_{\gamma\gamma}^2 - \widetilde{c}_{ZZ}^2 + \widetilde{c}_{\gamma\gamma}^2 + 2 c_Z \kappa_{ZZ} + 3 \left( c_{ZZ} \kappa_{ZZ} + c_{Z\gamma} \kappa_{Z\gamma} \right) \\
				&- \frac{c_w^2 - s_w^2}{s_w c_w} \left(c_Z \kappa_{Z\gamma} + c_{ZZ} c_{Z\gamma} + c_{Z\gamma} c_{\gamma\gamma} - \widetilde{c}_{ZZ} \widetilde{c}_{Z\gamma} - \widetilde{c}_{Z\gamma} \widetilde{c}_{\gamma\gamma}\right) \\
				&- \frac{c_w^2 - s_w^2}{s_w c_w} \frac{3}{2} \left(c_{Z\gamma} \kappa_{ZZ} + c_{\gamma\gamma} \kappa_{Z\gamma} + \kappa_{ZZ} \kappa_{Z\gamma}\right) + \frac{13 \kappa_{ZZ}^2 + 4 \kappa_{Z\gamma}^2}{3}
			\end{aligned}
			\right] \\
			&\; - 2 s_w^2 c_w^2 \frac{\ln{\widetilde{\Lambda}^2}}{16 \pi^2 v^2} \left[
			\begin{aligned}
				&\frac{2 m_Z^2}{3} \epsilon(Z) + m_Z^2 \left(c_{ZZ}^2 - c_{Z\gamma}^2 - \widetilde{c}_{ZZ}^2 + \widetilde{c}_{Z\gamma}^2\right) \\
				&+ m_H^2 \left[ c_{ZZ} \kappa_{ZZ} + c_{Z\gamma} \kappa_{Z\gamma} + 2 \frac{\kappa_{ZZ}^2 + \kappa_{Z\gamma}^2}{3} - \frac{c_w^2 - s_w^2}{2 s_w c_w} \left(c_{Z\gamma} \kappa_{ZZ} + c_{\gamma\gamma} \kappa_{Z\gamma}\right) \right] \\
				&+ \frac{c_w^2 - s_w^2}{s_w c_w} m_Z^2 \left[ c_Z \left(2 c_{Z\gamma} + 3 \kappa_{Z\gamma}\right) - c_{ZZ} c_{Z\gamma} + \widetilde{c}_{ZZ} \widetilde{c}_{Z\gamma} - c_{Z\gamma} \kappa_{ZZ} - \frac{3 \kappa_{ZZ} \kappa_{Z\gamma}}{2} \right]
			\end{aligned}
			\right]
		\end{split} \, ,
		\\[0.5cm]
		\begin{split}
		\alpha T =
			&\; \frac{3}{8} \frac{\Lambda^4}{16 \pi^2 v^2} \left[\frac{\kappa_{ZZ}^2 + \kappa_{Z\gamma}^2}{m_Z^2}-\frac{\kappa_{WW}^2}{m_W^2}\right] + f(\Lambda^2, 0, 0) \left[
			\begin{aligned}
				& c_Z^2 - c_W^2 + 3 c_Z \kappa_{ZZ} - 3 c_W \kappa_{WW} - 3 \frac{\kappa_{ZZ}^2 - \kappa_{WW}^2}{4} \\
				&- \frac{3 m_H^2}{4} \left(\frac{\kappa_{ZZ}^2 + \kappa_{Z\gamma}^2}{m_Z^2}-\frac{\kappa_{WW}^2}{m_W^2}\right)
			\end{aligned}
			\right] \\
			&+ 3 \frac{\ln{\widetilde{\Lambda}^2}}{16 \pi^2 v^2} \left[ m_Z^2 \left(c_Z^2 - 1\right) - m_W^2 \left(c_W^2 - 1\right) - m_Z^2 c_Z \kappa_{ZZ} + m_W^2 c_W \kappa_{WW} + \frac{m_Z^2 \kappa_{ZZ}^2 - m_W^2 \kappa_{WW}^2}{4} \right]
		\end{split} \, ,
		\\[0.5cm]
		\begin{split}
		\alpha U =
			&\; 2 s_w^2 f(\Lambda^2, 0, 0) \left[
			\begin{aligned}
				& c_{WW}^2 - \widetilde{c}_{WW}^2 + 3 c_{WW} \kappa_{WW} - 3 c_w^2 \left(c_{Z\gamma} \kappa_{Z\gamma} + c_{ZZ} \kappa_{ZZ}\right) - 3 c_w s_w \left(c_{\gamma\gamma} \kappa_{Z\gamma} + \kappa_{ZZ} \kappa_{Z\gamma} + c_{Z\gamma} \kappa_{ZZ}\right) \\
				&+ 2 \left(c_W \kappa_{WW} - c_w^2 c_Z \kappa_{ZZ} - c_w s_w c_Z \kappa_{Z\gamma}\right) + 13 \frac{\kappa_{WW}^2 - c_w^2 \kappa_{ZZ}^2}{3} - \frac{4 c_w^2 \kappa_{Z\gamma}^2}{3} \\
				&- \left(c_w c_{ZZ} + s_w c_{Z\gamma}\right)^2 - \left(c_w c_{Z\gamma} + s_w c_{\gamma\gamma}\right)^2 + \left(c_w \widetilde{c}_{ZZ} + s_w \widetilde{c}_{Z\gamma}\right)^2 + \left(c_w \widetilde{c}_{Z\gamma} + s_w \widetilde{c}_{\gamma\gamma}\right)^2
			\end{aligned}
			\right] \\
			&\; - \frac{2}{3} s_w^2 \frac{\ln{\widetilde{\Lambda}^2}}{16 \pi^2 v^2} \left[
			\begin{aligned}
				& 2 m_W^2 \epsilon(W) - 2 c_w^2 m_Z^2 \epsilon(Z) + 3 m_W^2 \left(c_{WW}^2 - \widetilde{c}_{WW}^2\right) + 2 m_H^2 \left(\kappa_{WW}^2 - c_w^2 \kappa_{ZZ}^2 - c_w^2 \kappa_{Z\gamma}^2\right) \\
				&+ 3 m_H^2 \left[ c_{WW} \kappa_{WW} - c_w \kappa_{ZZ} \left(c_w c_{ZZ} + s_w c_{Z\gamma}\right) - c_w \kappa_{Z\gamma} \left(c_w c_{Z\gamma} + s_w c_{\gamma\gamma}\right) \right] \\
				&- 3 m_Z^2 \left[ \left(c_w c_{ZZ} + s_w c_{Z\gamma}\right)^2 - \left(c_w \widetilde{c}_{ZZ} + s_w \widetilde{c}_{Z\gamma}\right)^2 - c_w s_w \left(2 c_Z - \kappa_{ZZ}\right)\left(2 c_{Z\gamma} + 3 \kappa_{Z\gamma}\right) \right]
			\end{aligned}
			\right]
		\end{split} \, .
	\end{align}
	\endresizemath\end{adjustwidth}

	\par After removal of the power divergences in the $S$, $T$, $U$ parameters by using relations \ref{eq:ki}, \ref{eq:cV}, \ref{eq:cWWcWWt} and \ref{eq:cZZcZZt}, only the following logarithmic contributions remain present:

	\vspace{-0.5cm}
	\resizemath{\small}
	\begin{align}
		\begin{split}
		\alpha S &= - s_w^2 c_w^2 \frac{\ln{\widetilde{\Lambda}^2}}{8 \pi^2 v^2} m_Z^2 \left[
			\begin{aligned}
				& c_{ZZ}^2 - \widetilde{c}_{ZZ}^2 - c_{Z\gamma}^2 + \widetilde{c}_{Z\gamma}^2 - \frac{c_w^2 - s_w^2}{s_w c_w}\left(c_{ZZ} c_{Z\gamma} - \widetilde{c}_{ZZ} \widetilde{c}_{Z\gamma}\right) \\
				&+ 2 \left(\frac{c_V^2 - 1}{3} - c_V c_{ZZ} - c_V c_{\gamma\gamma}\right)
			\end{aligned}
			\right] \\
			&= - s_w^2 c_w^2 \frac{\ln{\widetilde{\Lambda}^2}}{8 \pi^2 v^2} m_Z^2 \left[
			\begin{aligned}
				& c_{\gamma\gamma}^2 - \widetilde{c}_{\gamma\gamma}^2 - c_{Z\gamma}^2 + \widetilde{c}_{Z\gamma}^2 + \frac{c_w^2 - s_w^2}{s_w c_w}\left(c_{\gamma\gamma} c_{Z\gamma} - \widetilde{c}_{\gamma\gamma} \widetilde{c}_{Z\gamma}\right) \\
				&+ 2 \left(\frac{c_V^2 - 1}{3} - 2 c_V c_{\gamma\gamma} - \frac{c_w^2 - s_w^2}{s_w c_w} c_V c_{Z\gamma}\right)
			\end{aligned}
			\right]
		\end{split} \, ,
		\\[0.5cm]
		\alpha T &= 3 \frac{\ln{\widetilde{\Lambda}^2}}{16 \pi^2 v^2} m_Z^2 s_w^2 \left(c_V^2 - 1\right) \, ,
		\quad
		\alpha U = 0 \, .
	\end{align}
	\endresizemath

\clearpage


	\providecommand{\href}[2]{#2}\begingroup\raggedright\endgroup

\end{document}